\RequirePackage[running]{lineno}

\documentclass[aps,prl,nofootinbib,twocolumn,showpacs,superscriptaddress,letterpaper,amsmath,amssymb]{revtex4}

\pdfoutput=1

\usepackage{graphicx}  % needed for figures
\usepackage{dcolumn}   % needed for some tables
\usepackage{bm}        % for math
\usepackage{amssymb}   % for math
\usepackage{feynmf}%%%{feynmp}
\usepackage{slashed}
\usepackage{multirow}
\usepackage{subfig}
\usepackage{xcolor}
\usepackage[colorlinks=true,citecolor=blue,linkcolor=blue]{hyperref}

\def\gsim{\lower0.5ex\hbox{$\:\buildrel >\over\sim\:$}}
\def\lsim{\lower0.5ex\hbox{$\:\buildrel <\over\sim\:$}}

\newcommand{\be}{\begin{equation}}
\newcommand{\ee}{\end{equation}}
\newcommand{\bea}{\begin{eqnarray}}
\newcommand{\eea}{\end{eqnarray}}

\newcommand{\nbox}{{\,\lower0.9pt\vbox{\hrule \hbox{\vrule height 0.2 cm
\hskip 0.2 cm \vrule height 0.2 cm}\hrule}\,}}

\def\eg{{\it e.g.}}
\def\etc{{\it etc}}

\def\sub#1{_{\lower.25ex\hbox{$\scriptstyle#1$}}}

\newskip\zatskip \zatskip=0pt plus0pt minus0pt
\def\matth{\mathsurround=0pt}
\def\lsim{\mathrel{\mathpalette\atversim<}}
\def\gsim{\mathrel{\mathpalette\atversim>}}
\def\sigv{\ifmmode \langle\sigma v\rangle\else $\langle\sigma v\rangle$\fi}
\newskip\zatskip \zatskip=0pt plus0pt minus0pt
\def\matth{\mathsurround=0pt}
\def\lsim{\mathrel{\mathpalette\atversim<}}
\def\gsim{\mathrel{\mathpalette\atversim>}}
\def\atversim#1#2{\lower0.7ex\vbox{\baselineskip\zatskip\lineskip\zatskip
  \lineskiplimit
  0pt\ialign{$\matth#1\hfil##\hfil$\crcr#2\crcr\sim\crcr}}}

\begin{document}

\thispagestyle{empty}
\vspace*{-3.5cm}

\vspace{0.5in}

\newcommand{\kcnote}[1]{\textcolor{red}{[KC: #1]}}

%\begin{flushright}
%\today\\
%\end{flushright}
%\vspace{0.5in}
\title{Modeling Smooth Backgrounds \& Generic Localized Signals with Gaussian Processes}

\begin{center}
\begin{abstract}
We describe a procedure for constructing a model of a smooth data spectrum using Gaussian processes rather than the historical parametric description.  This approach considers a fuller space of possible functions, is robust at increasing luminosity, and allows us to incorporate our understanding of the underlying physics. We demonstrate the application of this approach to  modeling the background to searches for dijet resonances at the Large Hadron Collider and describe how the approach can be used in the search for generic localized signals.
\end{abstract}
\end{center}

\author{Meghan Frate}
\affiliation{Department of Physics and Astronomy, University of
  California, Irvine, CA 92697}
\author{Kyle Cranmer}
\affiliation{Department of Physics and Astronomy, New York University, New York, NY 10003}
\author{Saarik Kalia}
\affiliation{Department of Physics, MIT, Boston, MA}
\author{Alexander Vandenberg-Rodes}
\affiliation{Obsidian Security Inc., Newport Beach, CA 92660}
\author{Daniel Whiteson}
\affiliation{Department of Physics and Astronomy, University of
  California, Irvine, CA 92697}

\maketitle

\section{Introduction}

The search for new particles and interactions is a central focus of the research program of the Large Hadron Collider (LHC). Typically, such a search is cast in the language of a hypothesis test of a background model predicted by the standard model of particle physics. In some cases, the alternative hypothesis is specified by a particular theory or class of theories, in which case a practical task of the experimentalist is to identify a good discriminating variable and to construct models of the background-only and signal-plus-background hypotheses. This approach includes searches for resonances on a smooth background spectrum as well as non-resonant signals with well-specified signal characteristics. In addition, some searches for new physics aim at model-independence by making only weak assumptions about the deviation from the background. For instance, such a search might assume only that the signal manifests itself as a localized deviation without completely specifying the distribution.  A critical element in both approaches is a proper description of the smooth background spectrum. 

In an ideal case, the background intensity $f(x)$ as a function of observable $x$ would be known exactly. In this scenario, we can derive an expected total number of events $\nu \equiv \int f(x) dx$ and a probability density $p(x) \equiv f(x) / \nu$, which leads to the familiar unbinned extended maximum likelihood
%
%\footnote{In statistics literature, this type of probability model would be referred to as an (inhomogeneous) Poisson point process with an intensity $f(x)$. }
%
 for a dataset $\mathcal{D} =\{x_1, \dots, x_N\}$ with $N$ observed events
\begin{equation}\label{eq:pointProcess}
p( \mathcal{D}  ) = \text{Pois}(N | \nu) \;  \prod_{i=1}^N p(x_i ) \; .
\end{equation}

In realistic applications, we replace the ideal background prediction $f(x)$ with a more flexible background model. The traditional approach for searches with smooth background shapes  \cite{Alitti1991, ATLAS:2012ad, Abrahamyan:2011gv} is to introduce a family of functions $f(x|\theta)$ with some explicit functional form parametrized by $\theta$. 
The choice of functional form to describe the background is central to the new particle search, yet functional forms derived from first principles are almost never available. Instead, the typical approach is to select an ad-hoc parametric function with little-to-no grounding in the physics involved, but which fits reasonably well in collider data and simulated samples. As the luminosity of the collected datasets grow, however, the  discrepancies between the ad-hoc model and the true physical process are revealed. As the rigid form and limited flexibility of the parametric functions fail to accommodate the observed spectra, continual addition of new ad-hoc terms is required. In the coming period of high-luminosity collisions at the Large Hadron Collider, this issue will become even more acute. 

A second issue with the traditional approach is an unsatisfying bifurcation of the search strategy between hypothesis testing for specific signal models and the more model-independent search for localized deviations from a smooth background.  In model-specific hypothesis testing, a likelihood ratio test can be applied to weigh the background-only hypothesis against the signal-plus-background hypotheses, while the more model-independent searches often use a more elaborate procedure like {\sc BumpHunter}~\cite{Choudalakis:2011qn}. The {\sc BumpHunter} algorithm considers a large family of search windows and then performs a number-counting test in each. Both approaches face a look-elsewhere effect from multiple testing, but the {\sc BumpHunter}~\cite{Choudalakis:2011qn} approach also faces complications from the ad-hoc algorithmic choices and multiple correlated background estimates.

In this paper, we propose a new strategy for modeling smooth backgrounds, one that uses Gaussian Processes (GPs). This strategy relaxes the overly-rigid constraints of ad-hoc functional forms, which makes them more robust to increasing luminosity. 
%GPs can incorporate our understanding of the underlying physics. 
Moreover, GPs can be used to model generic localized signals, which allows for a unified statistical treatment of model-specific and more model-independent searches.

GPs are remarkably flexible and have been used to improve modeling for geostatistics, climate, exoplanets, and machine learning~\cite{sampson1992nonparametric, meiring1997developments, 0004-637X-806-2-215,rasmussen2006gaussian,grosse2012exploiting, DuvLloGroetal13}.  
It is curious that GPs have not yet been widely used in high-energy physics. We suspect that this is due to a few barriers, which we aim to address. 
First, the GP formalism is somewhat foreign to high-energy physicists, so we detail how it is that we can translate our understanding of the underlying physics into the GP formalism. Secondly, a large portion of the GP literature is focused on regression problems, instead of modeling probability distributions. Thus, we make an effort to connect the traditional extended likelihood treatment in high-energy physics to the GP approach. Lastly, there are subtle issues related to the statistical interpretation and procedures of GPs, so in addition to the more pedagogical goals of the paper we also incorporate some more technical statistical considerations for completeness. These technicalities are not essential to the main point, and should not deter the interested reader.

We use as a test case the LHC search for resonances decaying to two hadronic jets, which is one of the most powerful searches for new physics at the LHC and examines a falling spectrum over a broad mass range, providing a vigorous test of the method. In the following, we describe the historical approach, give the details of the GP approach and present studies that documents its performance. 
%The reader interested only in the practical details of GPs applied to the dijet example is invited to skip to the section on Performance Studies.

\section{Parametrized Function Approach}

\subsection{Dijet Resonance Search Example}

\begin{figure}
%\vspace{2cm}
%\centering
    \includegraphics[height=0.4\textwidth, angle=-90]{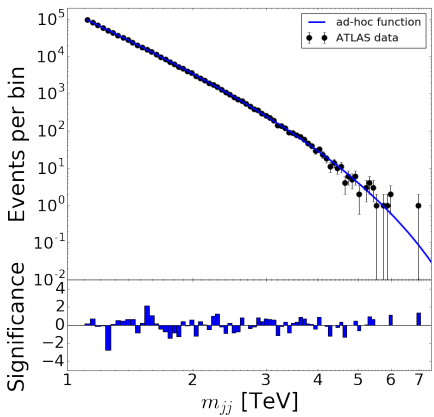}
\caption{Invariant mass of dijet pairs reported by ATLAS~\cite{ATLAS:2015nsi} in proton-proton collisions at $\sqrt{s}=13$ TeV with integrated luminosity of 3.6 fb$^{-1}$. The blue line is a fit using the first three terms of Eq.~\ref{eq:adhoc}. The bottom pane shows the significance of the residual between the data and the fit.}
\label{fig:funcdata}
\end{figure}

Searching for a resonance above a smooth background is a powerful technique in the hunt for new particles and interactions. This classic approach has been used in many experiments at different facilities, spanning several decades~\cite{Alitti1991,Shen:2009vs,Aaltonen:2008dn,Abazov:2009ac,Khachatryan:2010jd}. In particular, the search for dijets at the LHC is a simple but effective search for new physics. It analyzes the invariant mass spectrum of pairs of high-$p_{\textrm{T}}$ jets which are found back-to-back in the detector.  The dominant background is due to multi-jet production via non-resonant hadronic interactions, which historically has been difficult to model accurately  using simulation, from the bulk to the high-mass tail.  For this reason, a functional approach with three parameters was introduced by the UA2 collaboration~\cite{Alitti1991}. In the intervening thirty years, this form has needed the addition of further terms in order to be able to adequately describe observed spectra, leading to the most recent form used by the ATLAS collaboration~\cite{ATLAS:2015nsi} (previously $\theta_3=0$):
\begin{equation}
 f(x|\mathbf{\theta}) = \theta_0(1-x)^{\theta_1}x^{\theta_2}x^{\theta_3 \log x} 
\label{eq:adhoc}
\end{equation}
%\kcnote{original from paper... has 4 parameters not 5}
%\begin{equation}
% f(z|\mathbf{p}) = p_1(1-z)^{p_2}z^{p_3}z^{p_4 \log z} 
%\label{eq:adhoc}
%\end{equation}
%
%
%\noindent
where $x=m_{jj}/\sqrt{s}$. The selection of the particular functional form has been performed by comparing a handful of choices of varying complexity, selecting the simplest that satisfies the Wilk's test~\cite{wilks} which seeks ``to determine if the background estimation would be significantly improved by an additional degree of freedom''~\cite{ATLAS:2015nsi}. The structure of this parametric function and the selection of the terms is nearly entirely pragmatic, and has little basis in knowledge of the underlying physics processes.  An example fit of the three-parameter version to the ATLAS 2015 dijet data at $\sqrt{s}=13$ TeV is shown in Fig.~\ref{fig:funcdata}.

In some cases, experimental papers attempt to assess a systematic uncertainty on the result due to the arbitrary structure of this function~\cite{ATLAS:2015nsi}.   An alternative approach is to abandon the full-spectrum fit entirely and fit only a narrow sliding window~\cite{Abrahamyan:2011gv, Aaboud:2017yvp}, which tolerates a simpler functional form, but sacrifices the power of the full spectrum and complicates the global statistical interpretation. 

\subsection{Incorporating auxiliary information}

In the dijet case, the parameters $\theta$ are only constrained by the data $\mathcal{D}$; but more generally some auxiliary information $\mathbf{a}$ (\eg\ calibration measurements, theoretical considerations, \etc) may be used to constrain the parameters through an additional constraint term leading to the likelihood%
\footnote{In statistics literature, probability models of the form of Eq.~\ref{eq:pointProcess} are referred to as a Poisson point process with an intensity $f(x)$. }
\begin{equation}\label{eq:poissonPoint}
p( \mathcal{D}, \mathbf{a} | \theta ) = \text{Pois}(N | \nu(\theta)) \;  \prod_{e=1}^N p(x_e | \theta) \cdot p_\text{constr.}(\mathbf{a} | \theta) \; .
\end{equation}
In addition to constraint terms in the likelihood corresponding to auxiliary measurements $\mathbf{a}$ that have a clear frequentist interpretation, it is common to incorporate terms in the likelihood that reflect prior knowledge. For instance, uncertainties in theoretical cross section due to missing higher-order corrections typically are not treated as unconstrained parameters, instead the size of these uncertainties are estimated via varying the renormalization and factorization scales. Moreover, when very flexible models are used (\eg\ in unfolding problems) it is often desirable to include penalty terms that regularize the problem in order to avoid unphysical solutions. Lastly, consideration of multiple functional forms can fit into this formalism where $\theta$ takes on discrete values indexing the model choice~\cite{Dauncey:2014xga}. For a more pedagogical discussion of constraint terms, see Ref.~\cite{Cranmer:2015nia}.

\begin{figure*}[t]
%\vspace{4cm}
\includegraphics[width=.32\textwidth]{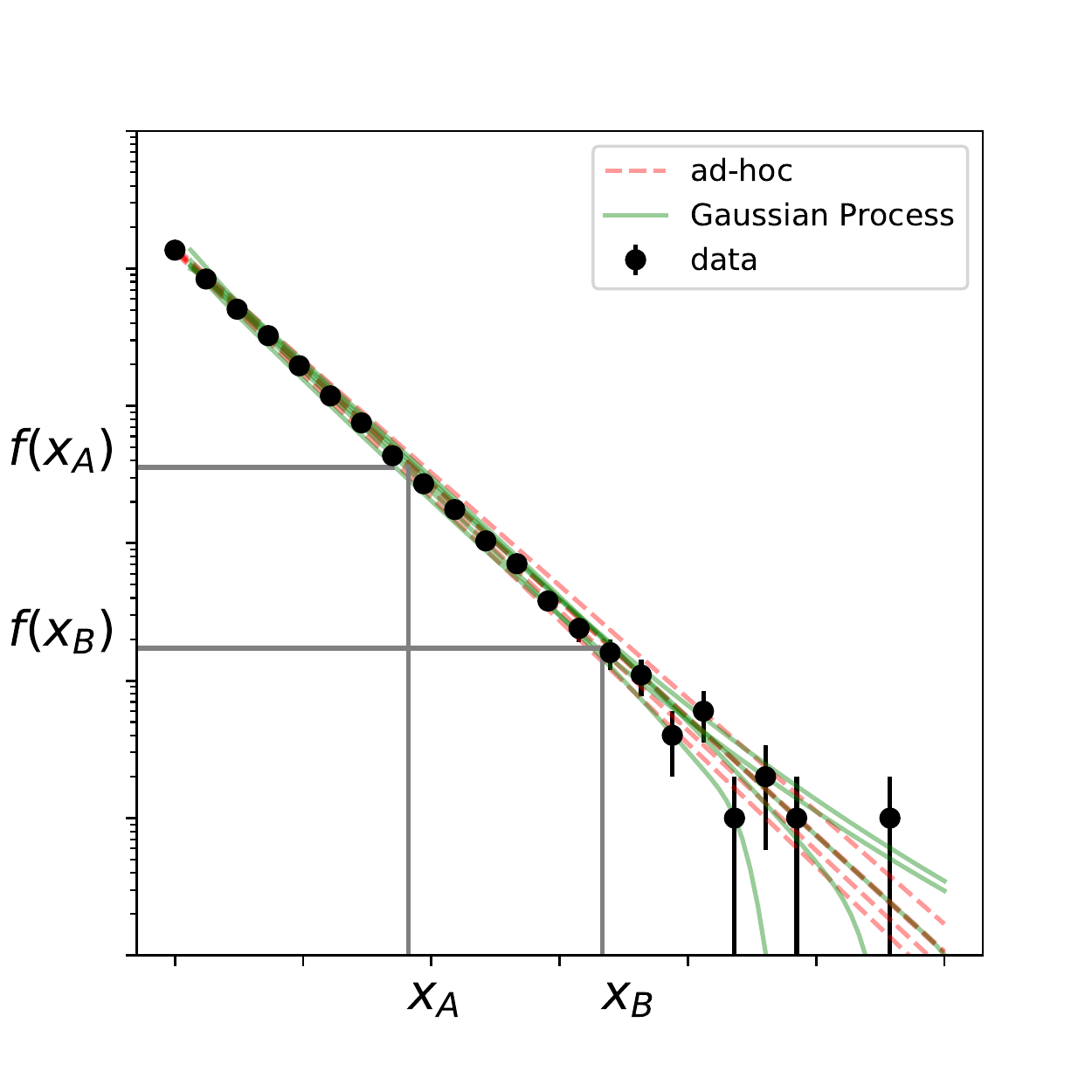}
\includegraphics[width=.32\textwidth]{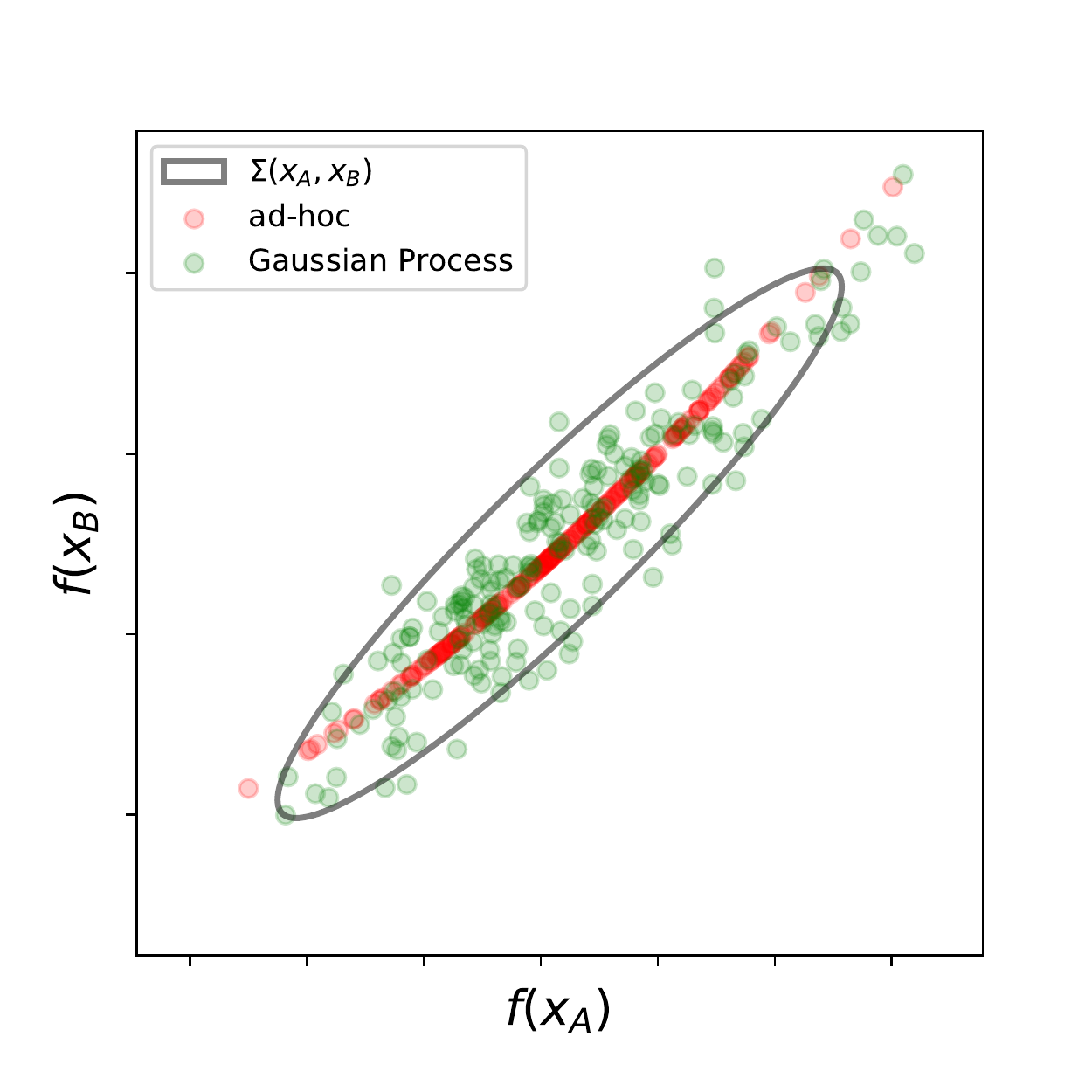}
\includegraphics[width=.32\textwidth]{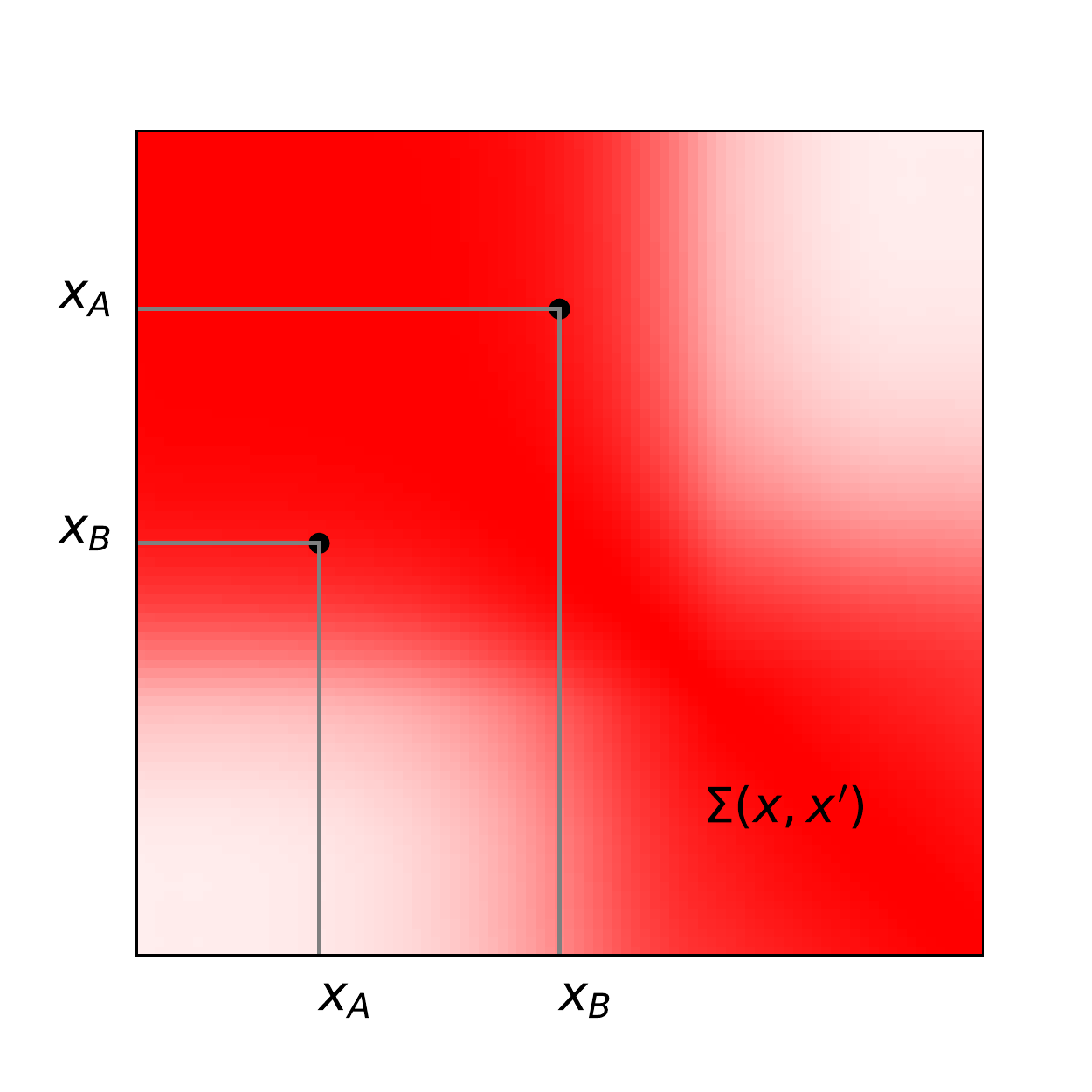}
\caption{Schematic of the relationship between an ad-hoc function and the GP. 
An example toy dataset is shown (left) with samples from the posterior for an ad-hoc 1-parameter function (red) and a GP (green). Each posterior sample is an entire curve $f(x)$, which corresponds to a particular point in the (center) plane of $f(x_A)$ vs. $f(x_B)$. The red dots for the ad-hoc 1-parameter function trace out a 1-dimensional curve, which reveals how the function is overly-rigid. In contrast, the green dots from the GP relax the assumptions and fill a correlated multivariate Gaussian (with covariance indicated by the black ellipse). The covariance kernel $\Sigma(x, x')$ for the GP is shown (right) with $\Sigma(x_A, x_B)$ corresponding to the black ellipse of the center panel.}
\label{fig:cartoon}
\end{figure*}

\section{Gaussian Process Approach}

Conceptually, GPs provide a generalization of the intensity $f(x)$ that is not tied to a particular functional form with a fixed number of parameters. Instead, the intensity at $x$ is modeled as a Gaussian and the relationship between the intensity at different points is encoded in the  \textit{covariance kernel} $\Sigma(x, x')$. In this way, GPs allow for the description of a much broader set of functions (see Fig.~\ref{fig:cartoon}) and provide a natural way to incorporate auxiliary information and  prior knowledge.

%It is curious that GPs have not yet been widely used in high-energy physics. We suspect that this is due to a few barriers. First, a large portion of the GP literature is focused on regression problems, instead of modeling probability distributions. Secondly, the GP formalism is somewhat foreign to high-energy physicists. In addition, there are some subtle statistical issues that arise in connecting GPs to traditional high-energy physics statistical procedures.
%
%In the remainder of this section, we provide a detailed discussion of the interpretation and statistical aspects of Gaussian processes in the context of high-energy physics. The reader interested only in the practical details of GPs applied to the dijet example is invited to skip to the section on Performance Studies.

%While the exact statistical treatment of Poisson statistical fluctuations with a Gaussian Process intensity is cumbersome\

While the exact treatment of Poisson statistical fluctuations together with GPs is somewhat cumbersome\footnote{From the point of view of the statistics literature, the combination of a Poisson point process with a Gaussian Process as the intensity is known as a Cox process, named after the statistician David Cox, who first published the model in 1955~\cite{coxprocess}. In order to enforce positivity in the intensity, it is common to model the log of the intensity  with a Gaussian process as was done in Ref.~\cite{golchi2015bayesian}. Inference in a doubly stochastic process is technically more cumbersome than the approach described above~\cite{Cunningham:2008:FGP:1390156.1390181}.}, in situations with many events we can simplify the situation by working with a binned likelihood where the Poisson counts in each bin $y_i$ are accurately approximated with a Gaussian distribution. In that case, the likelihood of Eq.~\ref{eq:poissonPoint} can be approximated as
\begin{eqnarray}
\label{eq:binnedPoisson}
p( \mathbf{y}, \mathbf{a} | \theta ) =&  \prod_{i=1}^n \text{Pois}(y_i | \bar{f}(x_i | \theta)) \cdot p_\text{constr.}(\mathbf{a} | \theta) \\ \nonumber
%&\approx& \prod_{i=1}^N \text{Gaus}(y_i |  f(x_i | \theta)) \times \text{Gaus}(f(\mathbf{x}) | \mathbf{\mu}, \mathbf{\Sigma}) \\
\label{eq:GPintro} \approx& \text{Gaus}(\mathbf{y} |  \bar{f}(\mathbf{x} | \theta), \mathbf{\sigma}^2) \cdot \text{Gaus}(\bar{f}(\mathbf{x|\theta}) | \mathbf{\mu}, \mathbf{\Sigma})  \, ,
\end{eqnarray}
where $\mathbf{x}$ are the bin centers, $\mathbf{y}$  are the observed bin counts, and $\bar{f}(\mathbf{x}|\theta)$ are the expected bin counts from averaging $f(x|\theta)$ within the corresponding bin. The first term of Eq.~\ref{eq:GPintro} is the per-bin Poisson statistical fluctuation, while the second term is an $n\times n$ multivariate Gaussian distribution that approximates the effect of $p_\text{constr.}(\mathbf{a}|\theta)$  propagated to the expected bin counts $\bar{f}(\mathbf{x}|\theta)$. 

The second term reveals how an ad-hoc functional form parametrized by $\theta \in \mathbb{R}^d$ can be overly rigid. In the $n$-dimensional space of bin counts, the parametrized function $\bar{f}(\mathbf{x} | \theta)$ only maps out a $d$-dimensional subspace (\eg~ the red markers in Fig.~\ref{fig:cartoon}).

Gaussian Processes provide a natural way to expand around the overly restricted parametrized model and fill in the full space of possibilities. A Gaussian Process is defined as ``a collection of random variables, any finite number of which have a joint Gaussian distribution'' \cite{rasmussen2006gaussian,golchi2015bayesian}. Instead of providing a parametric form, we directly model the mean 
($\mu$) and covariance functions ($\Sigma$)  of the Gaussian process defined as 
\begin{eqnarray}
\mu(x) &=& \mathbb{E}[f(x)] \\
\Sigma(x, x') &=& \mathbb{E}[(f(x)-\mu(x))(f(x')-\mu(x'))] \; .
\end{eqnarray}
This covariance \textit{kernel} $\Sigma$ is then augmented with the diagonal (uncorrelated) statistical component $\sigma^2(\textbf{x}) \mathbf{I}$ to provide the likelihood for the observed bin counts $\mathbf{y}$. 

\bigskip

%GPs are remarkably flexible and have been used to model atmospheric modeling, geostatistics, exoplanets, and machine learning~\cite{sampson1992nonparametric, meiring1997developments, 0004-637X-806-2-215,rasmussen2006gaussian,grosse2012exploiting, DuvLloGroetal13}.  
GPs have rarely been used in high-energy physics, with few exceptions including a top-quark mass measurement by the CMS experiment~\cite{Sirunyan:2017idq} and a paper by Golchi \& Lockhart focusing on other statistical aspects in the search for the Higgs boson~\cite{golchi2015bayesian}. Those analyses took advantage of the nice properties of GPs, but did not emphasize or explore how physics considerations come into play in the choice of the kernel. An attractive feature of GPs for high-energy physics is that the kernel directly encodes understanding of the underlying physics, which is manifest as covariance among the bin counts. 
For example, our understanding of the mass resolution, the parton density function uncertainties, the jet energy scale uncertainties, and expected smoothness can be incorporated into the kernel.

An advantage of interpreting the Gaussian process through the relationship of Eqs.~\ref{eq:binnedPoisson} and \ref{eq:GPintro} is
that it makes clear how to connect the GP formalism with the existing considerations around sources of uncertainty: those associated to auxiliary measurements with a clear frequentist interpretation, those associated to theoretical considerations that cannot be identified with a random variable with frequentist probability, and those terms that are introduced for the sake of regularization.

In the following sections, we outline the connection to unfolding, study the typical covariance structure from physical processes that affect the shape of the background spectrum, construct kernel and mean functions and describe the procedure to fit the GP to the observed spectrum.

\subsection{Connection to unfolding}

The process of constructing a GP kernel, which encodes our physical requirements for the background model, is closely related to the more familiar topic of imposing physical requirements when unfolding differential cross section distributions. 

In unfolding, we have an observed set of bin counts for an observable $x$, which we also assume to arise from a Poisson 
point process as in Eq.~\ref{eq:pointProcess}. In contrast to searches, the goal is to estimate 
the differential cross section for a theoretical quantity $z$, removing dependence on experimental efficiency, acceptance, and detector effects. 
The relationship between the target theoretical intensity $f(z)$ and the intensity for the observable  $f(x)$ is given by
\begin{equation}\label{eq:unfolding}
f(x) = \int f(z) W(x|z) dz \; ,
\end{equation}
where $W(x|z)$ is a folding matrix or transfer function that encodes smearing effects from detector resolution. 
Ideally, the experimentalist makes no assumptions about the theoretical intensity $f(z)$, and infers $f(z)$ through the unfolding process. 
As has been well studied, this type of inverse problem is ill-posed in the technical sense that the solution $f(z)$ is sensitive to small changes to $f(x)$ or observed data. The unregularized maximum likelihood solution to the unfolding problem often exhibits unphysical oscillations in $f(z)$.  Physical considerations motivate additional regularization or penalty terms to the likelihood that are not motivated by auxiliary measurements, but which lead to solutions that behave better for the $f(z)$ we consider relevant.~\footnote{There is a deep connection between Tikhonov regularization and other forms of regularization used in unfolding and the kernels of Gaussian Processes, which is formalized in the language of Reproducing Kernel Hilbert Spaces~\cite{rasmussen2006gaussian, PILLONETTO2014657}. A detailed discussion of this connection is beyond the scope of this paper; however, we should anticipate a contribution to the Gaussian Process kernel that is connected to this loosely defined notion of smoothness in the underlying physics.}

Of particular interest for particle physics is to revisit Eq.~~\ref{eq:unfolding} when $f(z)$ is itself a GP.  In that case Eq.~~\ref{eq:unfolding} can be seen as applying the linear operator $\int W(x|z) dz$ to the GP $f(z)$, which gives rise to another GP through a \textit{process convolution}~\cite{higdon1999non, paciorek2006spatial} resulting in
\begin{equation}\label{eq:smearingKernel}
\Sigma(x,x') = \int \int dz dz' \Sigma(z, z') W(x, z) W(x', z')
\end{equation}
As expected, even in the extreme case where different bins of the theoretical distribution are allowed to be totally uncorrelated, $\Sigma(z, z') \propto \delta(z-z')$, the finite resolution of the detector will introduce correlations in $x$ via $W$. For example, if $W(x|z)$ is a Gaussian smearing with resolution $\sigma_x$, then the resulting GP is an exponential squared kernel (see Eq.~\ref{eq:exp_sq_kernel}) with length scale $l=\sqrt{2} \sigma_x$.

\subsection{Physically motivated kernels}

Next, we motivate contributions to the kernel that are clearly grounded in experimental and theoretical forms of uncertainty cast in terms of Eq.~\ref{eq:unfolding}. In the ideal case, both the theoretical prediction for $f(z)$ and the efficiencies, acceptance, and experimental resolution encoded in $W(x|z)$ would be well specified. In that ideal case, $f(x)$ is totally fixed and the GP description of the intensity would collapse to a single point in function space. 

\begin{figure}
\centering
    \includegraphics[width=0.23\textwidth,trim={3.5cm 7.5cm 3.9cm 7.5cm},clip]{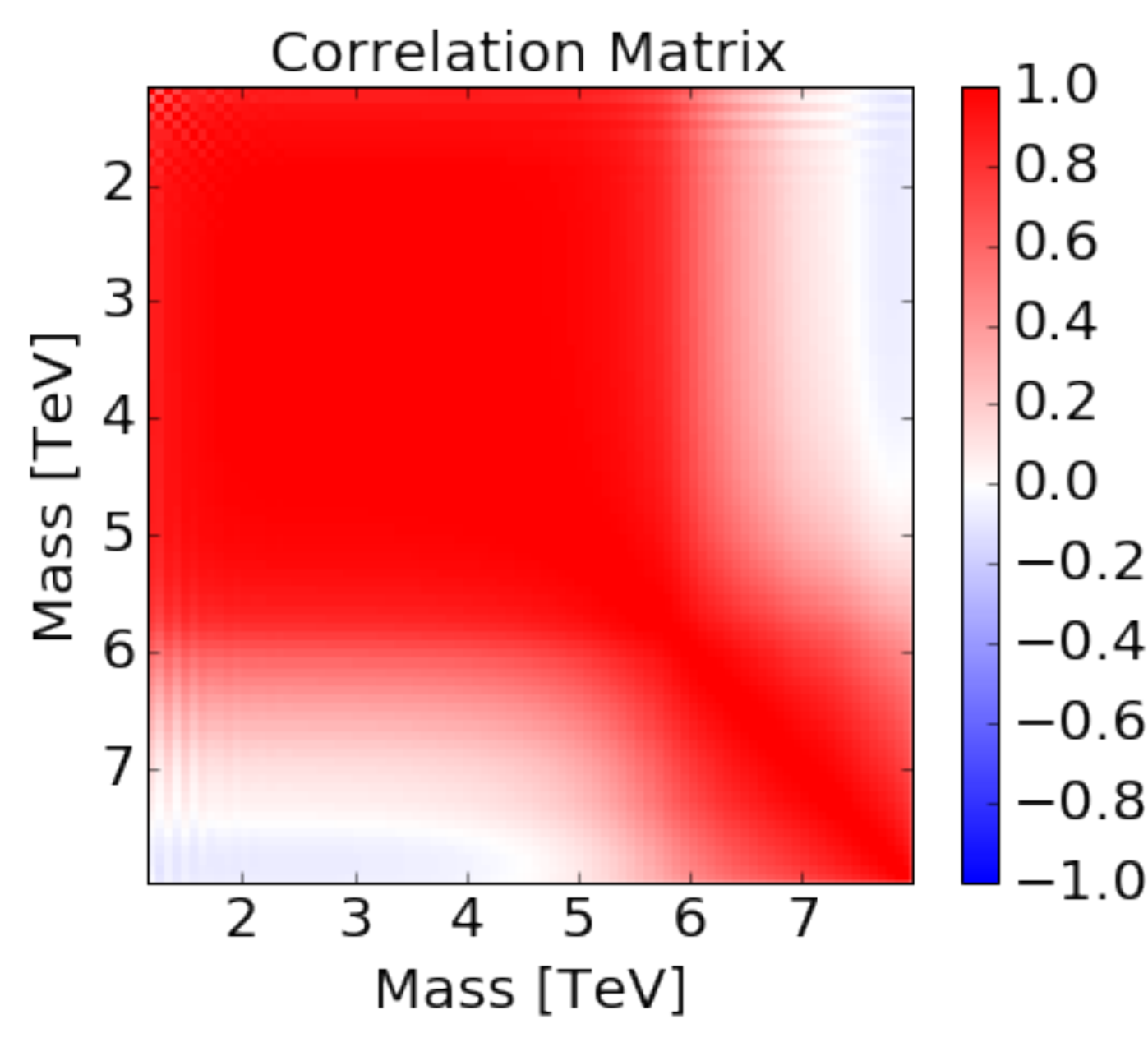}
    \includegraphics[width=0.23\textwidth,trim={3.8cm 7.5cm 3.6cm 7.5cm},clip]{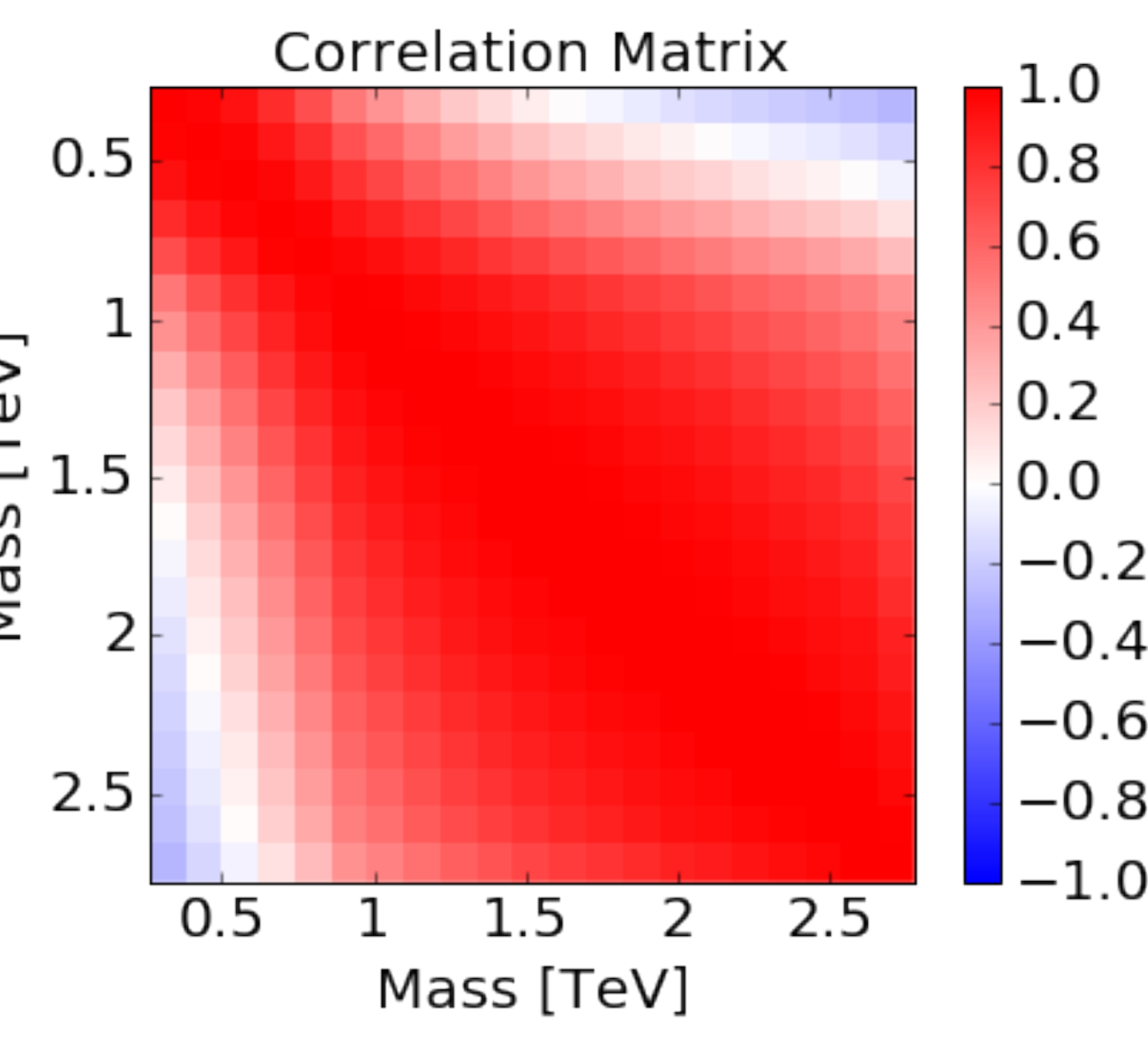}
\caption{ Correlation coefficients between pairs of mass bins due to variations in the jet energy scale (left) or parton distribution functions (right). These demonstrate the broad but smoothly varying influence of these effects on the mass spectrum.}
\label{fig:JESPDFCov}
\end{figure}

If, however, the detector response were itself uncertain, then this would propagate into the space of intensities. Take for example, the jet energy scale (JES) uncertainty. As described in Refs.~\cite{Aaboud:2017yvp,Aaboud:2016hwh} the ATLAS JES uncertainty is only a few percent for jets with $p_{\textrm{T}}$ of around 1 TeV where data are plentiful, while the the limited size of observed examples for higher-$p_{\textrm{T}}$ jets requires an alternate approach to estimating the JES. The resulting JES uncertainty therefore grows rapidly with $m_{jj}$ and has an impact of at most 15\%~\cite{Aaboud:2016hwh}. To illustrate the covariance due to the JES uncertainty, consider a simplified two-parameter model for the impact on the $m_{jj}$ distribution: $\textrm{J}(z, \theta) = 1+ 15\% \, \theta_1 z^4 + 5\% \, \theta_2 (1-z)$, where $z$ is the true dijet invariant mass and $z_\textrm{max}=7$ TeV. We use the best fit 3-parameter fit as a proxy for $f(z)$ and fold in the smearing 
 $W(x|z, \theta) = \textrm{Gaus}(x | z\, \textrm{J}(z/z_\textrm{max},\theta) , \sigma_x)$, where $\sigma_x=2\% z$ is the dijet invariant mass resolution~\cite{Aaboud:2017yvp}.  
 %Auxiliary measurements $\mathbf{a}$ can be used to estimate an uncertainty on the two components of the JES, which can be encoded into the constraint term $p_\text{constr.}(\mathbf{a}|\theta)$.  
 By assuming a uniform prior and an appropriate scaling for $\theta$, we sample from the posterior $\textrm{Gaus}(\theta_1 | 0, 1) \textrm{Gaus}(\theta_2 | 0, 1)$  and propagate the uncertainty in $\theta$ through to the predicted bin counts $\bar{f}(\mathbf{x} | \theta)$ as in Eqs.~\ref{eq:binnedPoisson} and \ref{eq:GPintro}. This allows us to explicitly build the covariance matrix $\Sigma$ using the simulation shown in Fig.~\ref{fig:JESPDFCov}. As expected, we see a roughly block-diagonal structure defined by low and high mass regions.

Similarly, we can study the uncertainty in the theoretical distribution arising from uncertainty in the parton distribution functions (PDFs), such as that in the ATLAS 7 TeV analysis~\cite{Aad:2013tea,alioli}. Figure~\ref{fig:JESPDFCov} corresponds to the PDF uncertainties described in Ref.~\cite{Alioli:2017jdo} for NLO calculations from \texttt{POWHEG-BOX}~\cite{Alioli:2010xd, Alioli:2010xa} 
using the $\mathcal{S}_\textrm{no-jet}$ PDF sets provided in \texttt{NNPDF3.0}~\cite{Ball:2014uwa}. In this case, sum rules in the PDFs lead to anti-correlation between low- and high-mass regions. 

With a satisfying GP description of the theoretical distribution and knowledge of the smearing $W(x|z)$, we can arrive at a well-motivated kernel for the observed spectrum via Eq.~\ref{eq:smearingKernel}.

\subsection{Implicit covariance in current background models}

\begin{figure}
\centering
    \includegraphics[width=0.23\textwidth,trim={3.8cm 7.5cm 3.9cm 7.5cm},clip]{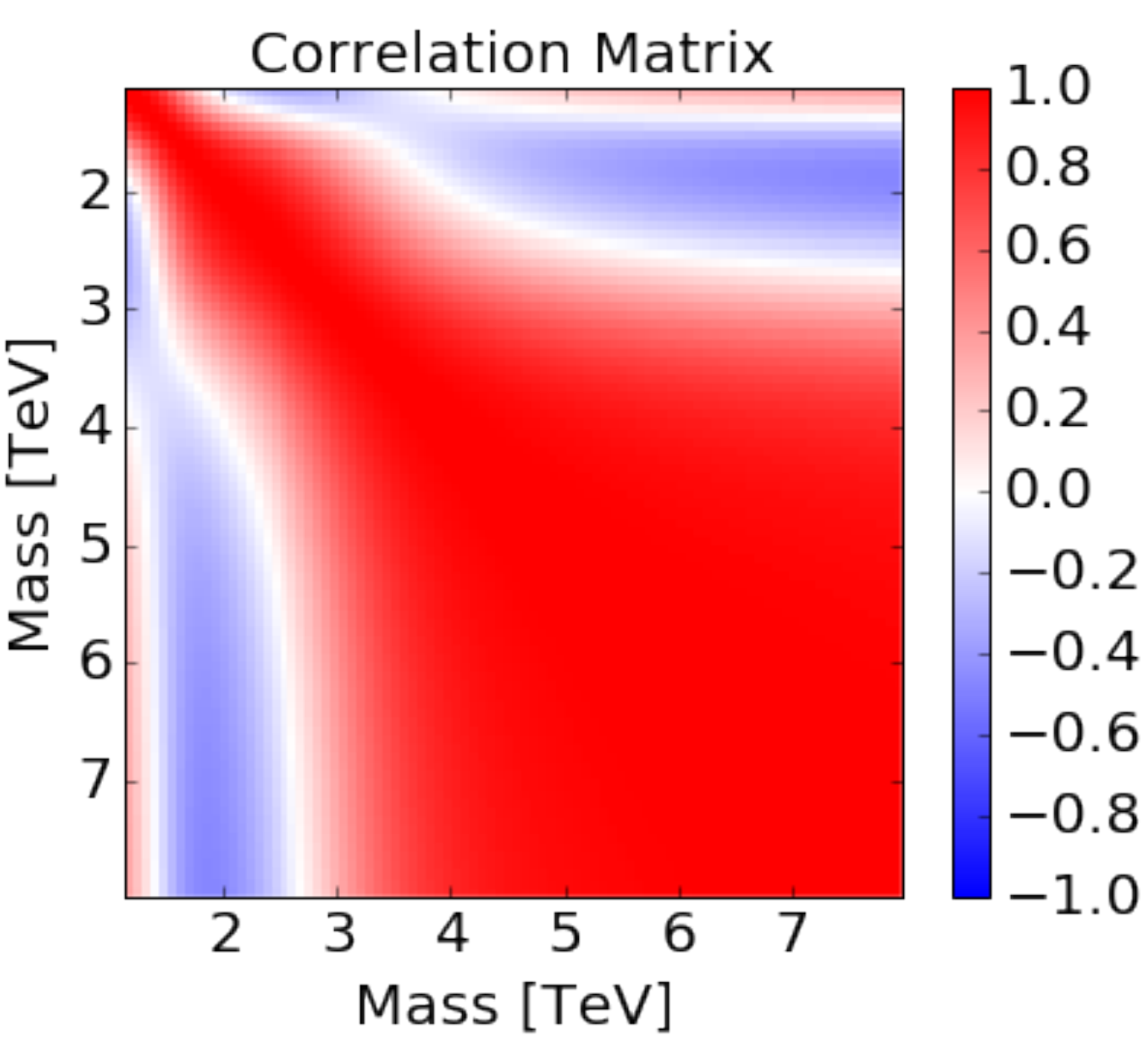}
    \includegraphics[width=0.23\textwidth,trim={3.8cm 7.5cm 3.9cm 7.5cm},clip]{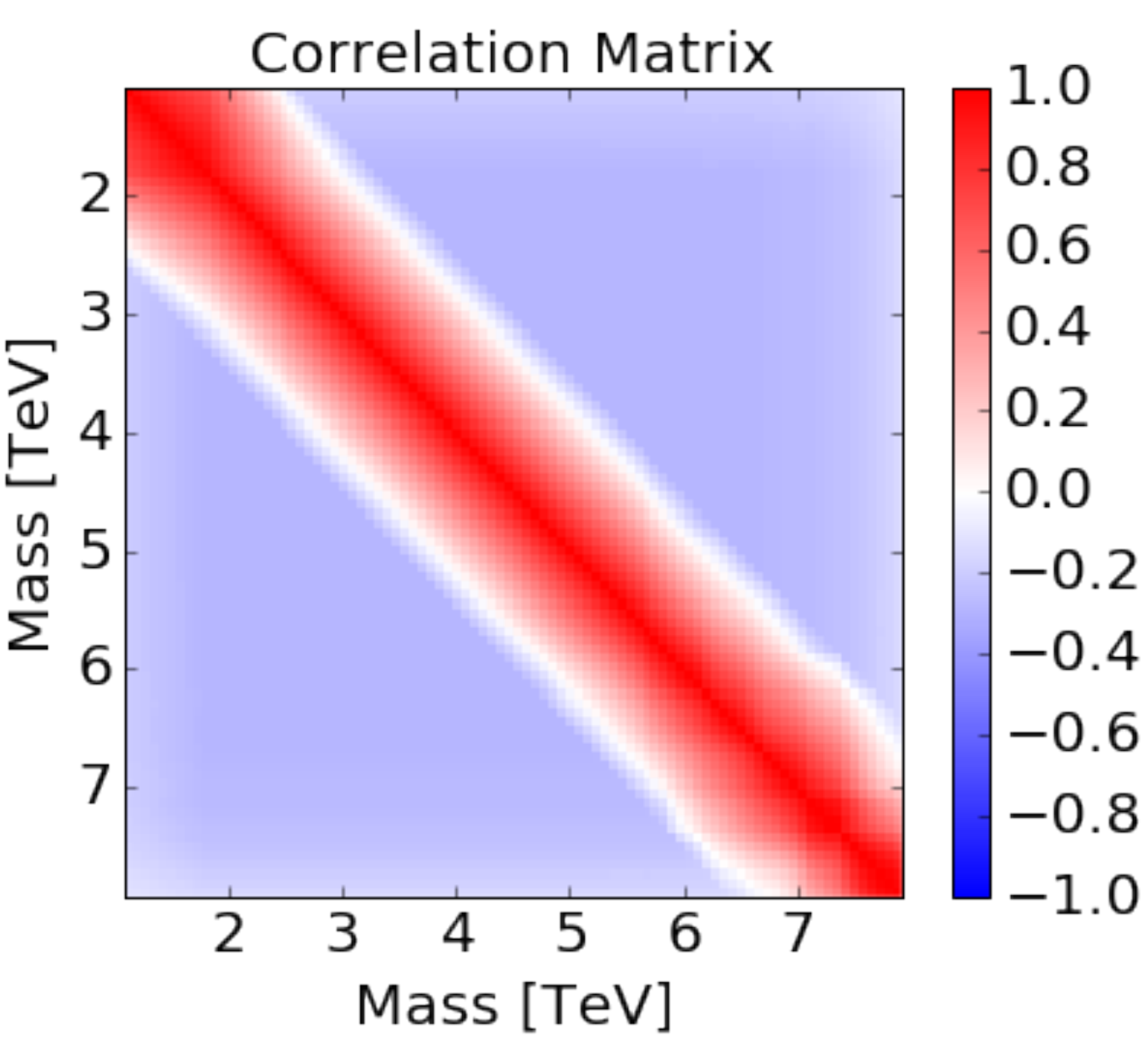}
\caption{Correlation coefficients between pairs of mass bins from many samples of the global ad-hoc fit (left) and the sliding window fit (right).  The plot of the global fit reveals non-physical pivot points where the ad-hoc function is less flexible. The sliding window fit has a strictly limited correlation, by construction.}
\label{fig:CompCov}
\end{figure}

It is also instructive to examine the effective covariance implied by current approaches: the ad-hoc  fit and the sliding window fit. Again we study this through the relationship of Eqs.~\ref{eq:binnedPoisson} and \ref{eq:GPintro}. In the case of the global fit to the ad-hoc function of Eq.~\ref{eq:adhoc}, we construct the posterior for $\theta$ given the ATLAS data shown in Fig.~\ref{fig:funcdata} and a uniform prior on $\theta$. We sample the posterior using \texttt{emcee}~\cite{emcee}, a Python implementation of the affine-invariant ensemble sampler for Markov Chain Monte Carlo (MCMC) proposed by Goodman \& Weare~\cite{goodman2010ensemble}. From the posterior samples of $\theta \sim p(\theta | \mathbf{y})$ we build the covariance matrix $\Sigma$ shown in  Fig.~\ref{fig:CompCov}. 
The global structure of the covariance resembles those arising from PDF uncertainties, but recall that the model only sweeps out a 3-dimensional subspace in the much larger space of functions with this same covariance.

In the case of the sliding window approach, we also estimate the covariance from a table of $f(x_i)$ values, but instead of posterior samples, we perform a single fit for each of 50 mass windows. For the $k^\textrm{th}$ window, if the bin is outside the window we record $0$, otherwise we record $f(x_i | \hat{\theta}_k)$, where $\hat{\theta}_k$ is the best fit value of $\theta$ for the fit restricted to the window. The covariance is calculated from these recorded values. This method should create a covariance structure which is limited to the diagonal band, as each fit includes only a small portion of the distribution -- indeed this is what we see%
\footnote{The slight negative correlation outside of this band is an artifact of recording $0$ for the value of the intensity, which is less than the mean.}
in Figure~\ref{fig:CompCov}.
While the  the sliding window fit approach provides more flexibility and better scaling to high luminosities than a global fit, the piecewise approach to background modeling complicates the downstream statistical analysis.

\subsection{Parametrized kernels}

In practice, one rarely works with a covariance matrix explicitly calculated from samples, but instead parametrizes the kernel $\Sigma(x, x')$. 
A common parametrized kernel is the exponential-squared kernel:
\begin{eqnarray}
\Sigma(x,x')= A \exp\left( {\frac{-(x-x')^{2}}{2 l^{2}}} \right) \; .
\label{eq:exp_sq_kernel}
\end{eqnarray}
This kernel describes a maximal covariance with amplitude $A$ that falls off with a length-scale $l$. If $|x-x'| \gg l$, then the covariance is very small. The parameters of the kernel are traditionally referred to as \textit{hyperparameters}, and the values of the hyperparameters can be fit either a priori (\eg\ by considering simulated or control samples), or to the data simultaneously with the hypothesis test itself.

In addition to the ability to construct kernels from first principles as in Eq.~\ref{eq:smearingKernel}, there are also rules for 
various types of operations on GPs~\cite{grosse2012exploiting, DuvLloGroetal13}.  For example, there are rules for describing the mean and covariance for a new GP produced from combining two GPs through the summation and multiplication. These rules effectively form a grammar that allows us to compose complex kernels through a narrative.

We can also build parametrized kernels that capture essential physics in a direct intuitive way. This approach sits somewhere between the first principles approach and the use of ad-hoc functions; however, unlike most ad-hoc functional forms, we will be able to interpret the terms more directly. For instance, in the dijet case we might argue that the mass resolution is one of the dominant contributions to the covariance, and start with an exponential squared kernel. We would anticipate the length scale to be larger than $\sqrt{2}\sigma_x$ from mass resolution effects to reflect the intrinsic smoothness of the underlying true distribution. In practice, the mass resolution is not constant, so we may allow for the length scale (incorporating both mass resolution and intrinsic smoothness) to have a linear dependence $l(x) = bx+c$. This varying length scale can be accommodated by the Gibbs kernel~\cite{gibbs1998bayesian, rasmussen2006gaussian}. In addition, instead of a constant amplitude for the variation, we take the amplitude of fluctuations to be a falling exponential. These considerations motivate the following parametrized kernel
%
%  In this kernel, the covariance between two points, $x$ and $x'$, includes a length scale function, in our case given by $l(x) = (bx+c)$. The definition of the length scale is critical, as it defines the smoothness of the GP description, which prevents the background model from accomodating the localized deviations from smoothness which are the resonance signals.  In this case, a length scale which increases with mass is used to describe the tightening smoothness requirements in the low-statistics tail, and to accomodate the dijet mass data from ATLAS, which has bins whose width increases with mass.  Finally, an amplitude term is used to model the variance with a falling exponential term. Our final kernel is:
%
\begin{eqnarray}
\Sigma(x,x')=&A e^{\frac{d-(x+x')}{2a}} %\\ \nonumber
  \sqrt{\frac{2l(x)l(x')}{l(x)^{2}+l(x')^{2}}} 
 e^{\frac{-(x-x')^{2}}{l(x)^{2}+l(x')^{2}}}
\label{eq:kernel}
\end{eqnarray}
The hyperparameters of this kernel $(A,a,b,c,d)$ will be determined during the fit to the data, described below. 
We can identify the terms associated to the narrative that produced this kernel, which we argue makes it less ad-hoc than
the functional form of Eq.~\ref{eq:adhoc}, and the GP associated to this kernel is much more flexible. 
Below we will compare the performance of this kernel to the ad-hoc function in the context of a dijet resonance search.

\subsection{Mean function}

It is common to use $\mu(x) =0$ for the mean of a GP, partially because once conditioned on the observations $\mathbf{y}$, the posterior mean usually adapts to this offset remarkably well if the spacing between the observed $x_i$ is small compared to the length scale. However, if we have a reasonable estimate of the mean, there is no reason not to use it. Thus we use the three-parameter fit functions as the mean $\mu(x)$ of the GP, which contributes an additional three hyperparameters ($\theta_{0}, \theta_{1}, \theta_{2}$). The results are very robust to the choice of the mean, the key to the performance is really in the choice of the kernel.

In the case of signal-plus-background hypothesis testing with a known signal model $f_s(x|\theta_s)$, the mean function  also includes the signal contribution. Due to the linear relationship between $\mathbf{y}$ and $\mu(\mathbf{x})$ in the Gaussian, this is numerically equivalent to subtracting the signal expectation from the observation and modeling the residuals with the background-only GP.

\subsection{Incorporating GPs into the statistical procedure}

There are subtle issues associated to the Bayesian and Frequentist interpretations of the GP. GPs are most commonly presented in a Bayesian formalism where the mean and covariance kernel are interpreted as a prior distribution over the space of functions. Then given the observations $\mathbf{y}$ we arrive at an updated GP that represents a posterior distribution over the space of functions. Because both the prior and the posterior are Gaussians there are explicit formulae for the posterior mean and covariance that rely on basic linear algebra~\cite{rasmussen2006gaussian}. In particular
\begin{equation}\label{eq:posteriorMean}
\mu(\textbf{x}_* |\mathbf{y}) = \mu(\textbf{x}_*) + \Sigma(\textbf{x}_*,\textbf{x})[\Sigma(\textbf{x},\textbf{x})+\sigma^{2}(\mathbf{x})\textbf{I}]^{-1}(\textbf{y}-\mu(\textbf{x})) 
\end{equation}
and
\begin{eqnarray}\label{eq:posteriorCov}
\Sigma(\mathbf{x}_*, \mathbf{x}_*') &=& \Sigma(\textbf{x}_*,\textbf{x}_*')  \\ \nonumber 
&-& \Sigma(\textbf{x}_*,\textbf{x}') [\Sigma(\textbf{x},\textbf{x}') + \sigma^2(\mathbf{x})\mathbf{I}]^{-1} \Sigma(\textbf{x},\textbf{x}_*') \;,
\end{eqnarray}
where $\textbf{x}_*$ are the values where the posterior GP is being evaluated and $\textbf{x}$ are the values being conditioned on.
In a typical binned analysis $\textbf{x}$ and $\textbf{x}_*$ would both be the the bin centers.
In addition, fitting the hyperparameters of a GP is usually based on maximizing the marginal likelihood, which has the explicit form
\begin{equation} %see eqtns 2.21-2.24 in R&M or 2.38
\log L = -\frac{1}{2} \log|\Sigma |-(\textbf{y}-\mu(\textbf{x}))^{T}\Sigma^{-1}(\textbf{y}-\mu(\textbf{x})) -\frac{n}{2} \log 2\pi \; .
\label{eq:marginallhood}
\end{equation}
%see also R&M Eq5.8

%In particular
%\begin{equation}\label{eq:posteriorMean}
%\mu(\textbf{x}|\mathbf{y}) = \mu(\textbf{x}) + \Sigma[\Sigma+\sigma^{2}(\textbf{x})\textbf{I}]^{-1}(\textbf{y}-\mu(\textbf{x})) 
%\end{equation}
%and
%\begin{equation}\label{eq:posteriorCov}
%\Sigma_\textrm{posterior}(\mathbf{x}, \mathbf{x}') = \Sigma - \Sigma[\Sigma + \sigma^2(\mathbf{x})\mathbf{I}]^{-1} \Sigma
%\end{equation}
%In addition, fitting the hyperparameters of a GP is usually based on maximizing the marginal likelihood, which has the explicit form
%\begin{equation} %see eqtns 2.21-2.24 in R&M or 2.38
%-\log L = \frac{1}{2} \log|\Sigma |+(\textbf{y}-\mu(\textbf{x}))^{T}\Sigma^{-1}(\textbf{y}-\mu(\textbf{x})) +\frac{n}{2} \log 2\pi \; .
%\label{eq:marginallhood}
%\end{equation}
%%see also R&M Eq5.8

\noindent In order to take advantage of the closed form solutions above and fast linear algebra implementations, the statistical fluctuations are typically approximated as $\sigma^{2}(\textbf{x}) = \textbf{y}$ instead of the more accurate Poisson mean.

In principle, one can revisit the logic of a particular GP kernel to trace back the terms that came from auxiliary measurements with a clear frequentist interpretation, and cary out the equivalent profile likelihood treatment. Given the correspondence between profiling and marginalization in the Gaussian case, this should lead to equivalent results with differences of at most constant factors; however, those factors may depend on the hyperparameters. The philosophical and practical use of the GP's profile likelihood will also be complicated by the contributions to the kernel from theoretical considerations that lack a frequentist interpretation and contributions motivated by regularization considerations. Thus, we leave this as a direction for future work. 

In practice there are roughly four ways the GP can be integrated into the  high-energy physics search strategy.

The first is to take on a fully Bayesian approach using the GP as the intensity of a Poisson point process, which forms a doubly stochastic Cox process~\cite{coxprocess}. Hypothesis testing and limits can then be based on Bayes factors. This approach was considered in Ref.~\cite{golchi2015bayesian}; however, it is computationally very intensive and difficult to combine with the bulk of the likelihood-based search strategies. 

The second is to approximate the Poisson fluctuations as Gaussian and use the marginal likelihood of Eq.~\ref{eq:marginallhood} directly in the test statistic, which is more computationally efficient. One can still use this marginal likelihood as a test statistic in the frequentist sense, but it requires using ensemble tests to calibrate the p-values as one cannot rely on the standard asymptotic formulae~\cite{Cowan:2010js}. 

The third approach is a two-step process where one first calculates the posterior mean of Eq.~\ref{eq:posteriorMean} and then uses $\mu(x)$ as the intensity for the Poisson likelihood of Eq.~\ref{eq:pointProcess} or its corresponding binned version. Similar to the profile likelihood approach, the background model can be conditioned on the signal hypothesis being tested, since the signal's parameters are present in the prior mean function. Computationally this approach is similar to the one above, with an additional cost of evaluating the Poisson likelihood and the practical considerations of managing the two-step process.

The last approach is also a two-step process where one first calculates both the posterior mean of Eq.~\ref{eq:posteriorMean} and the posterior covariance of Eq.~\ref{eq:posteriorCov} and then uses those in a down-stream least-squares analysis. For example, this approach is convenient for goodness-of-fit tests.

In the studies below we only use the marginal likelihood for fitting the hyperparameters, which we optimize using {\sc Minuit}~\cite{James:1975dr}. In later studies we use the third approach of treating the posterior mean (conditioned on the signal model parameters) as the Poisson intensity for the background.
We use the software package {\sc george}~\cite{hodlr} (see also \textsc{celerite}~\cite{celerite}) for GP regression, which we have extended by implementing custom kernels.

\begin{figure}
%\vspace{2cm}
%\centering
    \includegraphics[height=0.5\textwidth, angle=-90]{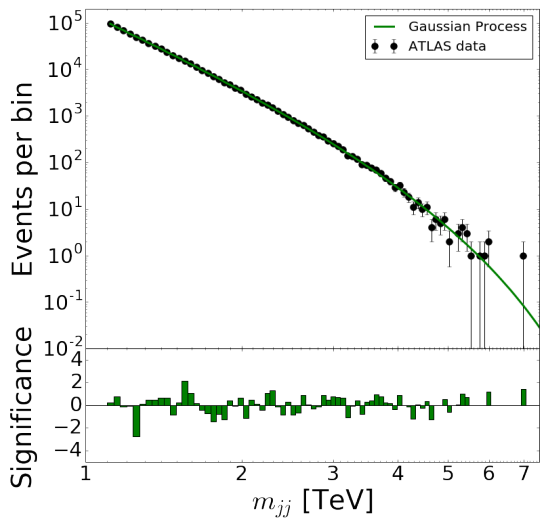}
\caption{Invariant mass of dijet pairs reported by ATLAS~\cite{ATLAS:2015nsi} in proton-proton collisions at $\sqrt{s}=13$ TeV with integrated luminosity of 3.6 fb$^{-1}$. The green line shows the resulting Gaussian process background model. The bottom pane shows the significance of the residual between the data and the GP model.}
\label{fig:gpdata}
\end{figure}

\begin{figure}
\centering
    \includegraphics[width=0.35\textwidth,trim={3.8cm 7.5cm 3.9cm 7.5cm},clip]{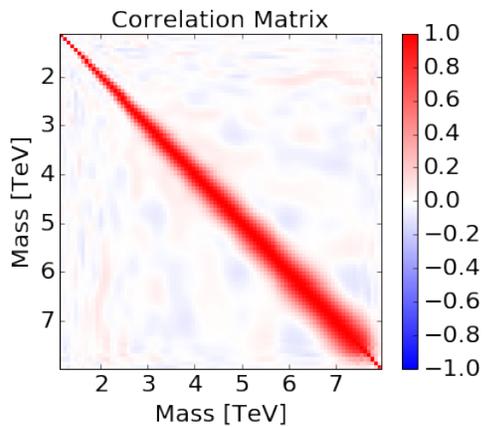}
\caption{Correlation between pairs of mass bins from the GP fit, which shows the largely diagonal nature, with increasing length scale at higher mass.}
\label{fig:gpcov}
\end{figure}

The posterior mean and posterior correlation matrix from fitting the GP to the ATLAS dataset are shown in Figs.~\ref{fig:gpdata} and \ref{fig:gpcov}. By visual inspection, the mean function fits the data well and the correlation is constrained near the diagonal, with the off diagonal dying off quickly. This structure reflects the locality of the GP, where nearby bins are closely connected but bins far from each other in mass are uncorrelated.

\section{Performance studies}

Figures~\ref{fig:gpdata} and \ref{fig:gpcov} demonstrate the fit (posterior mean) of the GP to a single dataset collected by ATLAS in proton-proton collisions at $\sqrt{s}=13$ TeV with integrated luminosity of 3.6 fb$^{-1}$~\cite{ATLAS:2015nsi}. More important is the characterization the GP approach from fits to an ensemble of datasets with independent statistical fluctuations and increasing luminosity.  We construct toy data samples by smoothing the ATLAS data, scaling it to the desired luminosity, and generating independent samples by adding Poisson noise to each bin.

Below, we present the performance of the GP approach in these datasets under two aspects of hypothesis testing:
\begin{itemize}
\item Background-only tests: these studies test whether the GP has sufficient flexibility to describe the typical background spectrum, assuming no signal.
\item Signal-plus-background tests: these studies combine a GP background with a specific signal model and tests the power of a hypothesis test based on the GP background. This requires that the GP model not be so flexible that it can absorb the localized signal into the background model.
%\item Anomalous signal search tests: these studies combine a GP background with a GP signal model, which allows experimenters to search for a non-specific localized signal.
\end{itemize}

\subsection{Background only tests}

\begin{figure}
%\vspace{2cm}
%\centering
    \includegraphics[height=0.5\textwidth, angle=-90]{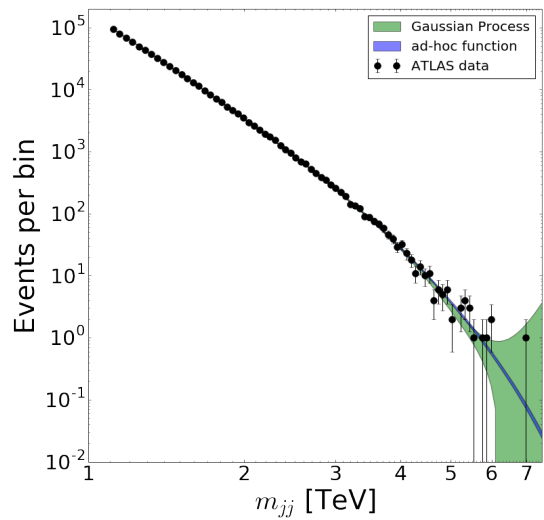}
\caption{ Tests of the Gaussian process and three-parameter ad-hoc function in toy data generated from the ATLAS data. Shown are the $\pm 1\sigma$ band about the mean background models, with the ATLAS data overlaid for reference.}
\label{fig:bgtoys}
\end{figure}

\begin{figure}
%\vspace{2cm}
%\centering
    \includegraphics[height=0.45\textwidth, angle=-90,trim={2.5cm 0cm 1.5cm 0cm},clip]{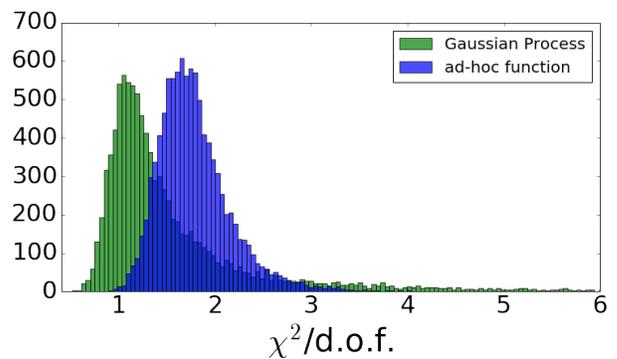}
\caption{The distribution of $\chi^{2}$ per degree of freedom in toy data generated from the ATLAS data at luminosity of 3.6 fb$^{-1}$. While the goodness of fit for the ad-hoc function degrades with more data, the GP is robust.}
\label{fig:chi2}
\end{figure}

The performance of the GP background model is evaluated in the toy datasets described above. For each toy dataset, we fit the GP, extract the posterior mean from Eq.~\ref{eq:posteriorMean}, and evaluate a $\chi^2$ quantity from $\sum_i (y_i - \mu(x_i|\mathbf{y}))^2/\mu(x_i|\mathbf{y})$; note that in this test we do not incorporate the posterior covariance matrix of Eq.~\ref{eq:posteriorCov}. 
Figure~\ref{fig:bgtoys} shows the $\pm 1\sigma$ about the average $\mu(\mathbf{x}|\mathbf{y})$ from these toys, with the ATLAS data to guide the eye, and the ad-hoc fit for comparison. The GP based on the kernel in Eq.~\ref{eq:kernel} has more flexibility at high mass, but also provides a superior fit, as measured by the $\chi^2$/dof statistic. The number of degrees of freedom for the ad-hoc fit is 3, while the GP has 8 hyperparameters (3 from the mean function and 5 from the kernel). Figure~\ref{fig:chi2} shows the distribution of $\chi^2$/dof, which peaks near $\chi^2$/dof $=1$ for the GP model and is significantly larger than unity for the ad-hoc function.

\begin{figure}
%%\vspace{2cm}
%\centering
    \includegraphics[height=0.45\textwidth, angle=-90,trim={3cm .3cm 1.5cm .5cm},clip]{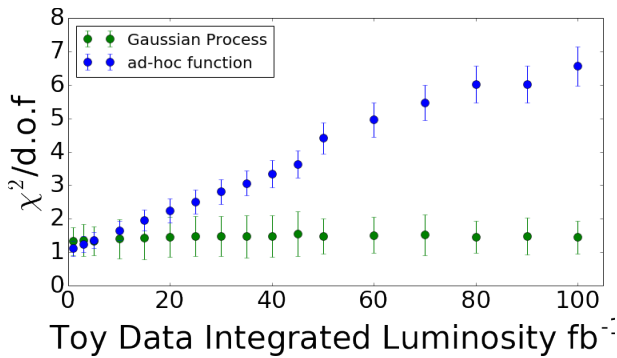}
\caption{ Mean and standard deviation of the $\chi^2$/d.o.f. measure in toy data generated from ATLAS collisions, as a function of integrated luminosity, for the ad-hoc fit and the Gaussian process.}
\label{fig:lumi}
\end{figure}

A critical test of the Gaussian process model is its robustness with increasing luminosity, where the ad-hoc approach has failed in collider data~\cite{ATLAS:2015nsi,Aaboud:2017yvp}. In Figure~\ref{fig:lumi}, the mean and standard deviation of the $\chi^2$/d.o.f. are shown as a function of integrated luminosity in the toy data, demonstrating the robustness of the GP approach.

\subsection{Background plus signal fits}

Adding more flexibility to the background model guarantees  a better fit to background-only toys; however, this generally comes at the loss of power in a search for a signal. A background model that is flexible enough to incorporate a signal contribution will have no discovery power.

Here, we test the GP model's performance in the toy data constructed as described earlier, but with signal injected as well.

\begin{figure}
%\vspace{2cm}
%\centering
    \includegraphics[width=0.4\textwidth, angle=-90]{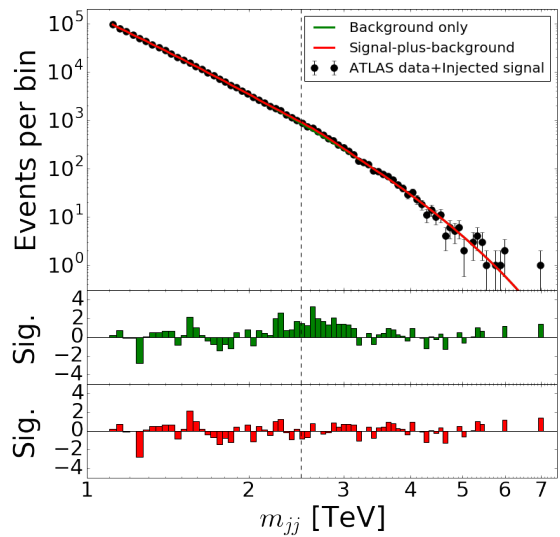}
\caption{Invariant mass of dijet pairs reported by ATLAS~\cite{ATLAS:2015nsi} in proton-proton collisions at $\sqrt{s}=13$ TeV with integrated luminosity of 3.6 fb$^{-1}$ with a false signal injected at $m_{jj}=2.5$ TeV. The green line is the Gaussian process background-only model; the red line is the signal-plus-background model. The central pane shows the significance of the residual between the data and the background fit; the bottom-pane shows the significance of the residual between the data and the background-plus-signal fit.}
\label{fig:sb1}
\end{figure}

We used a generic Gaussian resonance, and performed tests with various values for the signal mass, width and amplitude. The  hyperparameters of the GP (both for the background mean and kernel functions) are fixed from our fit to the ATLAS dataset; in a realistic application, experimenters could establish the hyperparameters in simulated samples.  We only fit the three parameters (amplitude, mass, and width) of the Gaussian signal. For the parametric fit, we fit all six parameters: the three fit function parameters and three signal parameters.  An example of this background-plus-signal fit is shown with an injected 2.5 TeV Gaussian signal shape in the top panel of Fig.~\ref{fig:sb1}.

This single example is illustrative and qualitative, but the statistical test for the presence of a signal in observed data relies on the  likelihood ratio $\Lambda$ between the background-only and the signal-plus-background hypotheses. We calculate the likelihood ratio between the two hypotheses in cases background-only toy data as well as background-plus-signal toy data.  This involves the use of Eq.~\ref{eq:posteriorMean} twice, as the posterior mean background prediction is different for the background-only and signal-plus-background fits. This is analogous to the profile likelihood ratio where there are two fits and the conditional maximum likelihood estimate of the background in the background-only case is generally different from the background estimate in the signal-plus-background fit.

\begin{figure}
%%\vspace{2cm}
%\centering
    \includegraphics[width=0.5\textwidth, angle=-0,trim={0cm 3cm 0cm 3cm},clip]{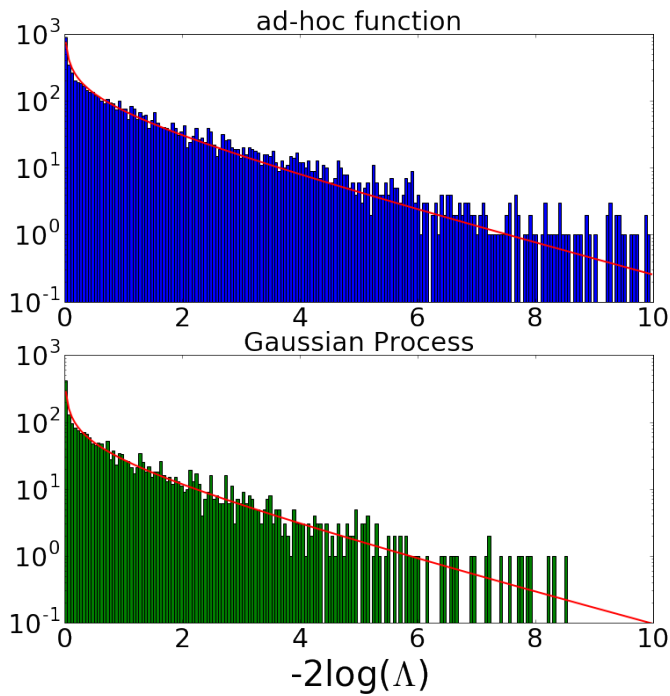}
\caption{Distribution of $-2 \log(\Lambda)$, where $\Lambda$ is the likelihood ratio between the background-only and the background-plus-signal hypotheses, for toy data with no signal present, shown for both the ad-hoc fit (top) and the Gaussian process background model (bottom). Overlaid in red is a $\chi^{2}$ distribution with one degree of freedom. }
\label{fig:wilks}
\end{figure}

The distribution of $-2\log \Lambda$ is shown in Figure~\ref{fig:wilks} for background-only toys for both the 
3-parameter ad-hoc function and the GP.  In these fits the signal mass and width were fixed and  the signal strength was treated as the parameter of interest. 
In the parametric case, we can invoke Wilks' theorem, which says this distribution should follow
a chi-square distribution if the true distribution generating the data corresponds to some point in the parameter
space of the background model~\cite{wilks}. However, in this case, the background-only toys were not generated from 
the ad-hoc function, instead they were generated from a smoothed version of the ATLAS data. 
Nevertheless, the distribution closely tracks a chi-square distribution. 

In the case of the GP, the situation
is more subtle because of the 2-step nature of the statistical approach and the subtle Bayesian vs. Frequentist issues.
Because of the Gaussian form, we expect correspondence between the posterior mean and the maximum likelihood estimate, thus we two-step nature is an irrelevant technical detail. The more subtle issue is that  the likelihood of Eq.~\ref{eq:pointProcess} only reflects the Poisson fluctuations, while the constraint terms the kernel encodes are not reflected in this likelihood. In this case there is not significant tension between the data and the covariance kernel so the likelihood ratio distribution also tracks a chi-square distribution. In general, this will need to be checked explicitly.

\begin{figure}
%%\vspace{2cm}
%\centering
    \includegraphics[height=0.45\textwidth, angle=-90,trim={1.5cm 0cm 1.5cm 0cm},clip]{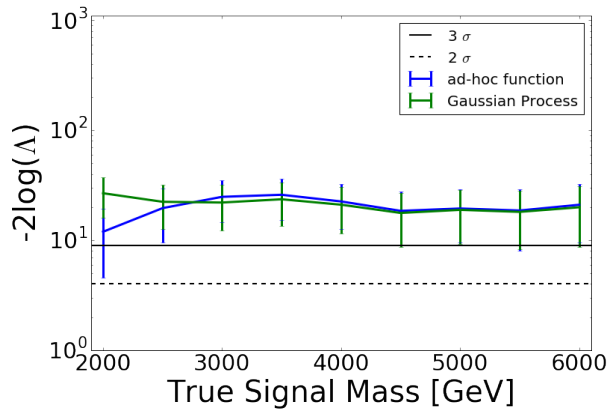}
\caption{ Mean log likelihood ratio ($\Lambda$) between the background-only and the background-plus-signal hypotheses, shown both for the 
of Gaussian process model and the ad-hoc  fit,  in 1000 toy data sets for varying injected signal mass. Solid and dashed lines indicate the threshold for $3\sigma$ significance and for an $\alpha$-level of 0.05.}
\label{fig:lhood}
\end{figure}

Next we directly assess the power of the search by considering the distribution of $-2\log\Lambda$ for signal-plus-background toys with signals of various masses.  Figure~\ref{fig:lhood} shows the mean of the $-2\log\Lambda$ distribution for the ad-hoc function and the GP model. The added flexibility of the GP does not degrade the power of the search -- in fact, the GP has more sensitivity to a signal at low mass, while the two methods are comparable at high mass. This gain in power is logically possible because the distribution used to true generate the background (which is normally unknown) does not correspond to the ad-hoc function exactly. In our case the GP background model is able to more accurately follow the true background (generated from smeared ATLAS data) than the ad-hoc function.

\begin{figure*}
%\vspace{2cm}
%\centering
\includegraphics[width=0.4\textwidth]{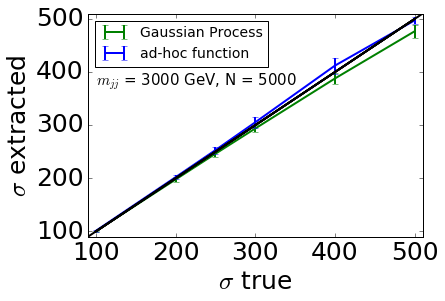}
\includegraphics[width=0.4\textwidth]{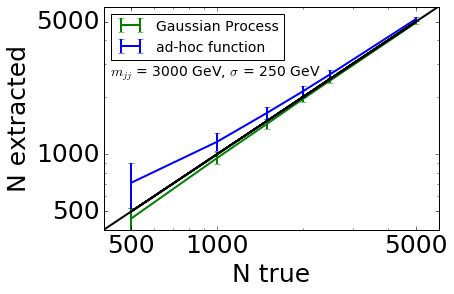}\\
{\includegraphics[width=0.4\textwidth]{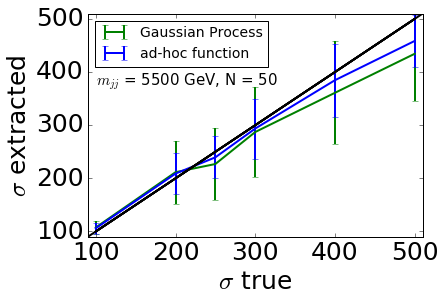}}
{\includegraphics[width=0.4\textwidth]{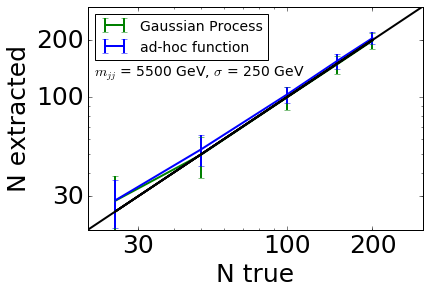}}
\caption{ Extracted signal parameters versus true parameters for an injected Gaussian signals with $m_{jj}=$ 3 TeV (top) and $m_{jj}=$ 5.5 TeV (bottom). Left: the extracted signal width ($\sigma$) for a fixed signal yield. Right: the extracted signal yield ($N$) for a fixed signal width ($\sigma=250$ GeV).  Results are shown for both the Gaussian Process background model (green) and the ad-hoc fit function (blue).
}
 \label{fig:varySig5}
\end{figure*}

If a signal is detected, it is also vital to be able to extract the signal parameters. For a two choices of signal mass ($m_{jj}=3$ and $5$ TeV), we performed fits to signal-plus-background toys fitting the mass, width, and signal strength. Figure ~\ref{fig:varySig5} shows that the extracted signal width and yield are reliable estimators of the true values.

%https://tex.stackexchange.com/questions/160730/putting-two-column-wide-figure-in-revtex4-1

\section{Modeling generic localized signals}

The search for specific resonances above a smooth background is only one type of search strategy.  More broadly, we hope to be sensitive to localized deviations that take different, potentially unanticipated shapes. For instance, a cascade decay can lead to triangular distributions with a sharp endpoint~\cite{Allanach:2002nj}, though helicity correlations can modify this shape in detail. Searches like this require balancing a small number of tests of the background-only model using generic properties of a signal and a larger number of tests of the background-only model using more specific signal properties. A single number-counting search using the full mass range is very generic, but has very little power. Conversely, an enormous scan over specific hypothesized signals individually have more power, but this strategy suffers from a large look-elsewhere effect.

Historically, the search for generic signals over a background model has used algorithms such as {\sc Bump-Hunter}~\cite{Choudalakis:2011qn}. This approach imposes only minimal structure on the signal: that it is a localized, contiguous excess. 
This  approach can be effective, but it has significant practical drawbacks as it contains many ad-hoc algorithmic elements. For example, in common usage  {\sc BumpHunter} requires that the excess be localized to at least two bins, and at most half of the bins. This algorithmic characterization of the signal is effective, but it is difficult to interpret and characterize statistically. Secondly, to address the look-elsewhere effect, this approach explicitly accounts for  multiple testing and calibrates the distribution of the test statistic by applying the entire procedure to background-only toys.   This requires a global background-only prediction, which is complicated when we rely on the data to help fit the background model. In particular, if a signal is present in the data, it is unclear how this impacts the background estimate. Thus far, the main strategy has been an iterative background estimation procedure that defines a signal region and extrapolates the background fit into this region. This approach introduces a coupling of algorithmic decisions with the statistical considerations. Similarly, in the context of a sliding window background model, the procedure is further complicated by the fact that there is not a single global background prediction, but a set of correlated background predictions specific to the signal window under consideration.

In this section, we consider an alternative approach which uses a GP to model a generic localized signal. In this case, the basic physical requirement of the localized signal can be encoded directly in the kernel of the signal GP, rather indirectly through ad-hoc algorithmic choices. This approach allows us to perform signal-plus-background fits where the signal GP absorbs the localized excess and the background GP accounts for the background. The background component from such a fit can provide global background estimate to be used in the context of a \textsc{BumpHunter} approach even when a signal is present in the data. More importantly, this approach enables hypothesis tests of the background-only model against a weakly specified signal-plus-background model directly based on the likelihood ratio or Bayes factor. 

We initiate the study of this approach with a specific signal GP described by the following kernel
\begin{equation}
\Sigma(x,x') = A e^{-\frac{1}{2}(x-x')^{2}/l^{2}} e^{-\frac{1}{2}((x-m)^{2}+(x'-m)^{2})/t^{2}}, 
\end{equation}
which has three main terms. The first term $A$ is an overall amplitude for the signal. The second term is the standard exponential-squared kernel with length scale $l$. The third term is an envelope that localizes the signal around a mass $m$ with a width $t$, which is analogous to the mass window.

\begin{figure}
%\vspace{2cm}
%\centering
    \includegraphics[width=0.4\textwidth,trim={0 0 0 0},angle=-90,clip]{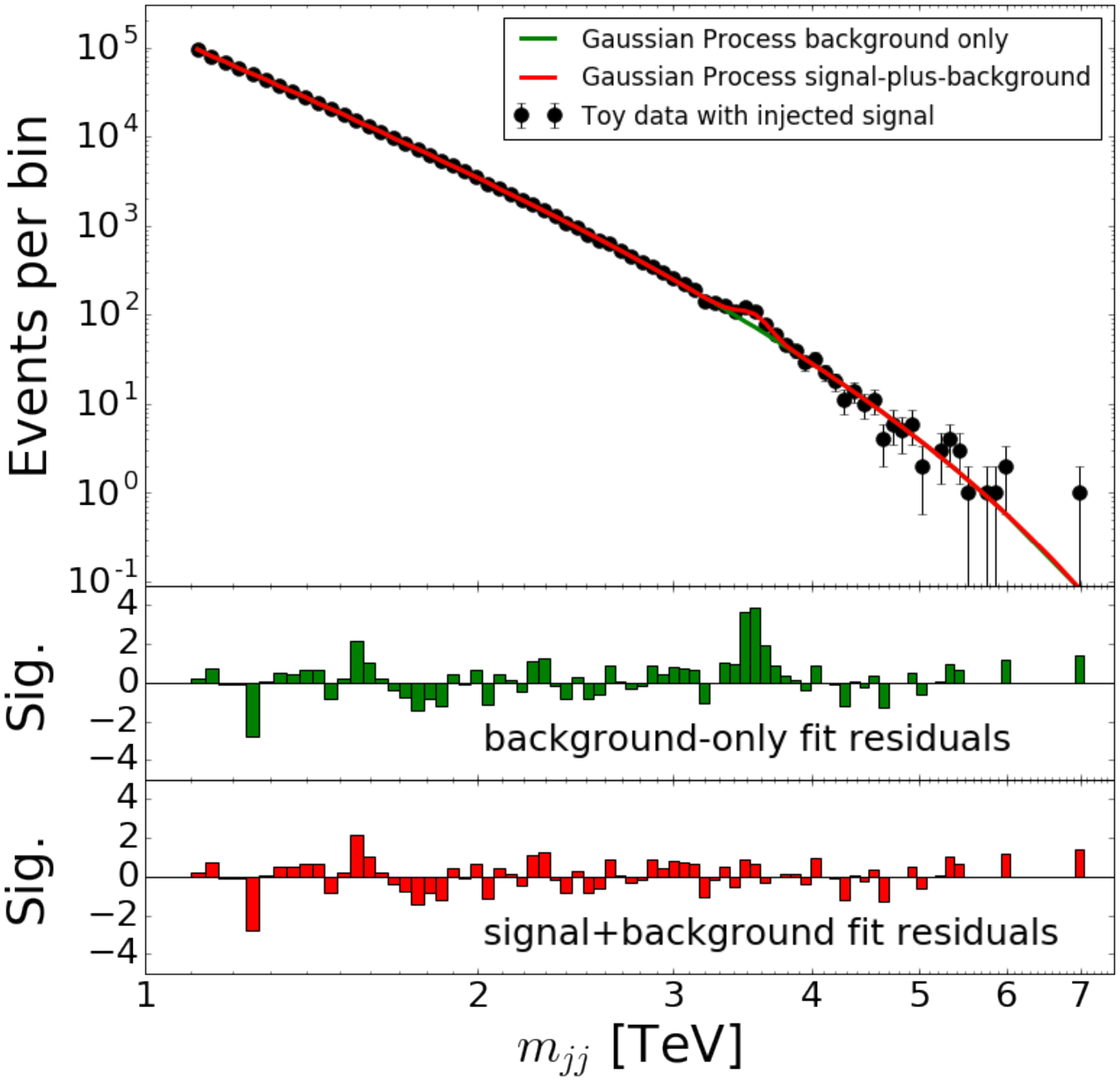}
\vspace{1cm}
    \includegraphics[width=0.4\textwidth,trim={.0cm .2cm 0.3cm 0},angle=-90,clip]{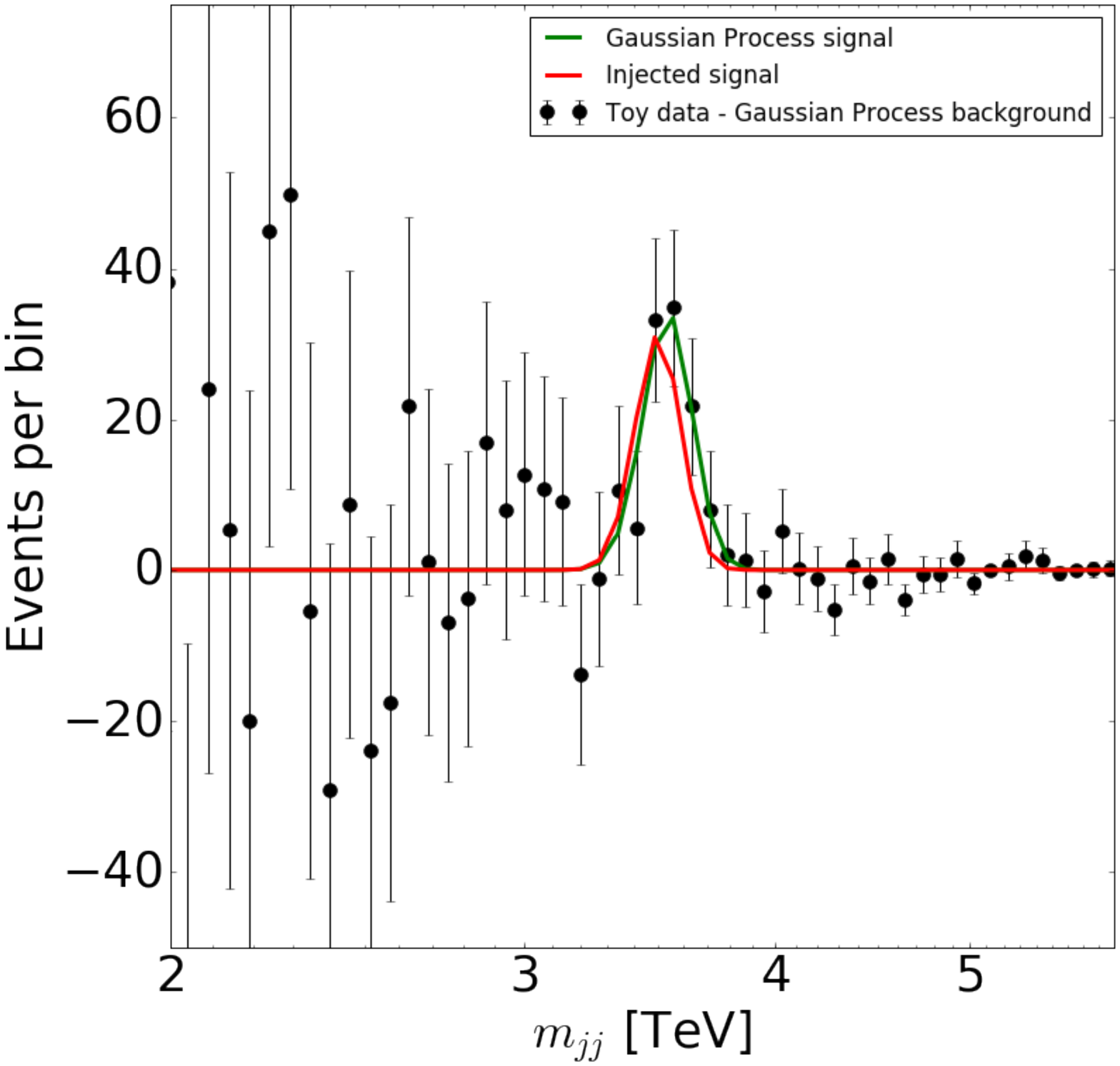}
\caption{ Top, an example fit with both a GP background and signal model to toy data with injected triangular signal.  The panes below show the significance of residuals between the toy data and the background model, the toy data and the background-and-signal model. Bottom, the residuals between the toy data and the background model, overlaid with the injected signal and the fitted GP signal.}
\label{fig:otherSignalShapes}
\end{figure}

\begin{figure}
%\vspace{2cm}
%\centering
        \includegraphics[width=0.4\textwidth,angle=-90]{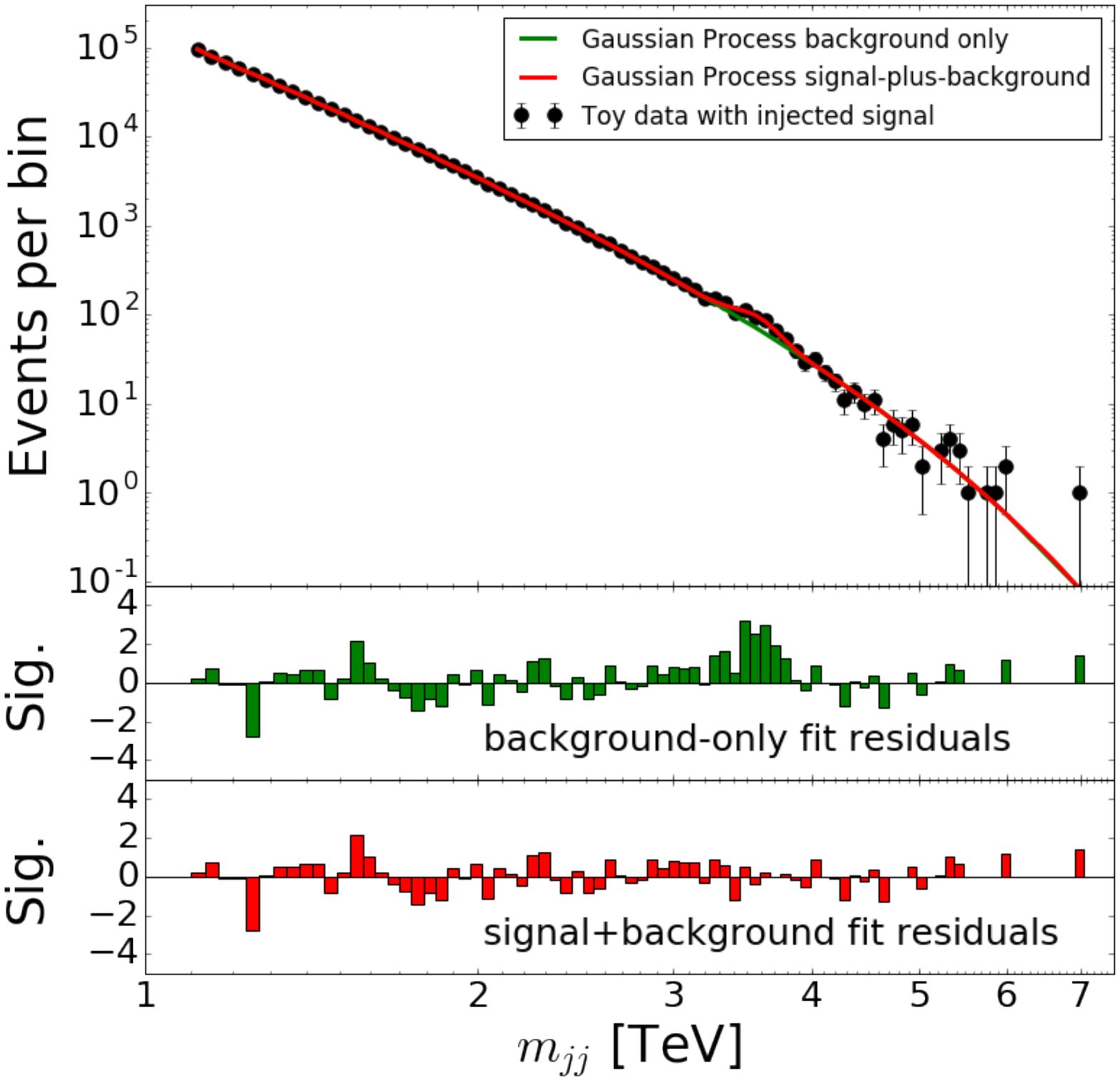}
\vspace{1cm}
        \includegraphics[width=0.4\textwidth,trim={0cm .2cm 0.3cm 0},angle=-90,clip]{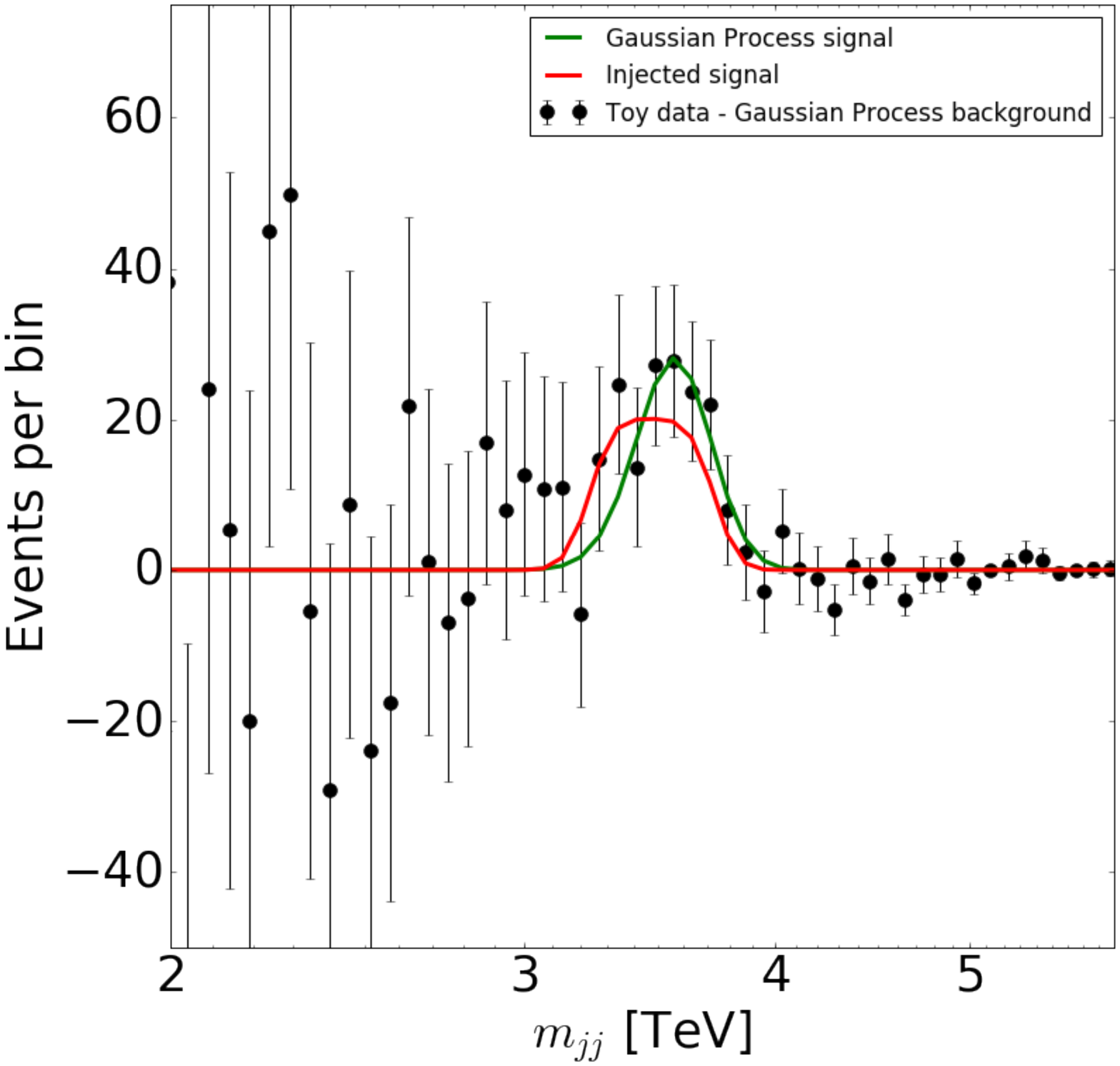}
\caption{ Top, an example fit with both a GP background and signal model to toy data with injected square signal.  The panes below show the significance of residuals between the toy data and the background model, the toy data and the background-and-signal model. Bottom, the residuals between the toy data and the background model, overlaid with the injected signal and the fitted GP signal.}
\label{fig:otherSignalShapes2}
\end{figure}

To demonstrate the flexibility of this kernel, we  performed signal-plus-background fits for a variety of signal shapes. Figure~\ref{fig:otherSignalShapes} shows the background extraction on both a linear piecewise triangular signal and Figure~\ref{fig:otherSignalShapes2} shows a square signal; both have been smeared to model detector jet energy resolution effects. Our studies indicate the GP signal is able to accommodate a wide variety of signal shapes leaving the background model responsible for for the smoother background-only component.

\subsection{Look-elsewhere effect}

This approach does not eliminate the look-elsewhere effect that arises from considering multiple signal hypotheses. Instead of a finite number of search windows or signal hypotheses, the GP describes a continuous family of signal hypotheses. This is not fundamentally different than the look-elsewhere effect that arises from considering a signal model with an unknown mass or width, though it is in a non-parametric setting. While both GP and the the simple example of an unknown mass correspond to an infinite number of signal hypotheses, they are highly correlated and the effective trials factor is finite~\cite{Gross:2010qma, Vitells:2011da}. 

Fundamentally, the fact that some parameters of the signal model (\eg\ mass and width) have no effect in the background only-case (in statistics jargon, they are \textit{not identified under the null}~\cite{doi:10.1093/biomet/74.1.33}) means that the conditions necessary for Wilks's theorem are not satisfied and the log likelihood ratio distribution will not take on the chi-square form. While there are approaches to estimate the asymptotic distribution of the likelihood ratio test statistic for signal models with one or a few parameters~\cite{Gross:2010qma, Vitells:2011da}, we are not aware of an asymptotic theory in the case of GPs. The lack of an asymptotic theory has little practical impact since even in the case of signal models with a few parameters, the asymptotic distributions are only accurate for very significant ($\gtrsim 4\sigma$) excesses, and background-only toys are usually used in the interesting region of $2-5\sigma$. 

\begin{figure}
%%\vspace{2cm}
%\centering
    \includegraphics[width=0.5\textwidth, angle=0,trim={0cm 0cm 0cm 0cm},clip]{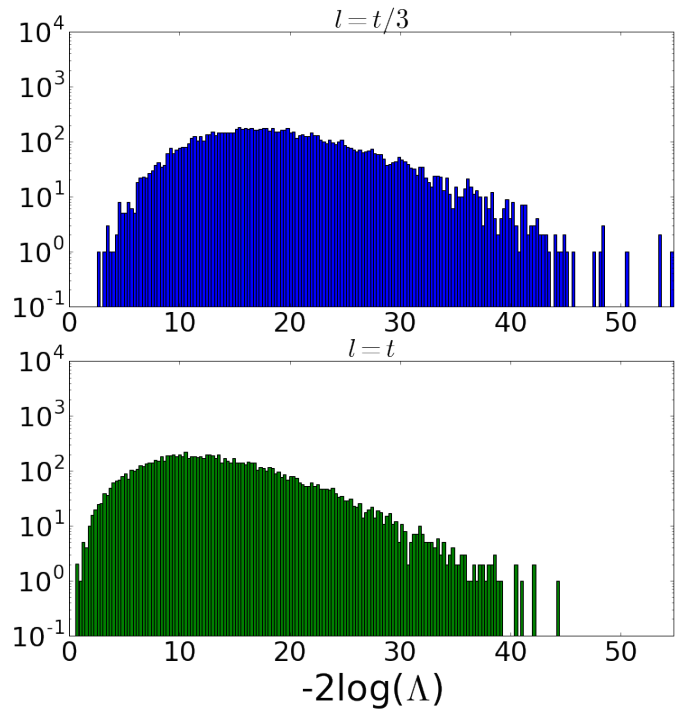}
\caption{Distribution of $-2 \log(\Lambda)$, where $\Lambda$ is the likelihood ratio between the background-only and the background-plus-signal hypotheses, for toy data with no signal present. The deviation from the $\chi^2_1$ distribution is due to the look-elsewhere effect. The top plot corresponds to a signal GP with $l=t/3$, which has more flexibility and a larger trials factor than the bottom plot with  $l = t$.}
\label{fig:LEE}
\end{figure}

The effective trials factor will depend on the specific background model and the kernel used for the signal GP. To illustrate this, we examined the log-likelihood ratio distribution for an ensemble of background-only toys similar to what was done in Fig.~\ref{fig:wilks}. In this case we fit the mass hyperparameter $m$ in the range 2-5 TeV and  fixed the hyperparameter $t=600$ GeV, which specifies that the signal is localized roughly to a 600 GeV region. Naively, the trials factor from allowing the mass to float (range over width) to be about 6. In addition we consider two different values for the length scale: $l=t$ and $l=t/3$. Smaller values for $l$ allow the signal GP more flexibility within the effective mass resolution, and thus further increase the trials factor. Figure~\ref{fig:LEE} shows the log-likelihood ratio distribution from these tests, confirms our intuition that smaller values of $l$ imply a larger look-elsewhere effect,  and demonstrates that it is straight forward to directly calculate the \textit{global} $p$-value from background-only toy Monte Carlo.

%Just as it is overly conservative to use a Bonferoni correction to the $p$-value based on in a binned ~\citep{Bonferroni36}.

\section{Discussion}

In this paper, we give a broad overview of the potential to use Gaussian Processes to model smooth backgrounds and generic localized signals. The use of GPs for modeling smooth backgrounds addresses the shortcomings of the current approaches based on ad-hoc parametrized functions. In particular, overly rigid parametrized functions lead to problems in high-luminosity searches because the inevitable deviations between the functional form and the true distribution can no longer hide behind statistical fluctuations. In contrast, the GP approach relaxes the rigid structure of a parametrized function, while maintaining the necessary notions of smoothness.

We have outlined the logical continuity between the GP likelihood to the conventional binned and unbinned Poisson likelihoods used in high-energy physics. We have discussed the interpretation of the kernel from first principles through a precise connection to unfolding and investigated the kernels associated to specific experimental and theoretical sources of uncertainty. 

We have studied in detail the performance of a simple intuitive kernel designed for dijet resonance searches. Finally, we have introduced a novel strategy for the search for generic localized signal excesses in which the weak specification of signal properties is provided via a GP kernel instead of the ad-hoc algorithmic choices of current approaches.

Gaussian Processes have improved the statistical modeling in a number of scientific fields, and these studies demonstrate their potential for high-energy physics. Their robustness in a high-luminosity setting and their ability to model weakly specified signals are both welcome and timely.

\section{Acknowledgements}

We would like to thank Simone Alioli and Duccio Pappadopulo for providing the parton density function uncertainties for the ATLAS dijet analyses. Duccio Pappadopulo, Lukas Heinrich, and Gilles Louppe provided useful feedback on the manuscript. We also thank Dan Foreman-Mackey for advice and support regarding custom kernels in \texttt{george}. DW and MF are supported by the Department of Energy Office of Science. Additionally, MF supported via NSF ACI-1450310. KC is supported through NSF ACI-1450310, PHY-1505463, PHY-1205376, and the Moore-Sloan Data Science Environment at NYU.

\bibliography{paper}

\clearpage
\appendix

\end{document}